\documentclass[10pt, journal , letterpaper]{IEEEtran}
\usepackage{amsmath,amssymb,amsfonts,amsthm}
\usepackage{algorithm}
\usepackage{algorithmic}
\usepackage{graphicx}
\usepackage{xcolor}
\usepackage{romannum}
\usepackage{setspace}
\usepackage[caption=false]{subfig}
\usepackage{tabularx}
\usepackage{pdfpages}
\allowdisplaybreaks
\graphicspath{{FigureCOMP/}}
\def\ie{{\it i.e.,\ \/}}

\DeclareMathOperator{\tr}{tr}
\DeclareMathOperator{\diag}{diag}
\DeclareMathOperator{\blkdiag}{blkdiag}
\theoremstyle{definition}
\newtheorem{thm}{Theorem}
\newtheorem{lem}[thm]{Lemma}
\newtheorem{remark}{Remark}

\makeatletter

\begin{document}

\pagenumbering{arabic}
\title
{\vspace{-2.5mm}\begin{Huge}Online Multi-Cell Coordinated MIMO Wireless Network Virtualization with Imperfect CSI\end{Huge}}
\author{
        Juncheng Wang, \IEEEmembership{Student Member, IEEE},
        Ben Liang, \IEEEmembership{Fellow, IEEE},\\
        Min Dong, \IEEEmembership{Senior Member, IEEE},       
        and Gary Boudreau, \IEEEmembership{Senior Member, IEEE}
\thanks{
J. Wang and B. Liang are with the University of Toronto (e-mail:
\{jcheng.wang, liang\}@ece.utoronto.ca). M. Dong is with the Ontario Tech
University (e-mail: min.dong@ontariotechu.ca). G. Boudreau is with Ericsson Canada (e-mail: gary.boudreau@ericsson.com).
This work has been funded in part by Ericsson Canada and by the Natural
Sciences and Engineering Research Council (NSERC) of Canada. A preliminary version of this work has appeared in IEEE INFOCOM \cite{INFOCOM20},
which studies only the single-cell case. This new version contains substantial
revision with multi-cell models, analysis, and simulation results.}
\vspace{-2.5mm}
}
\maketitle

\begin{abstract}
We consider online coordinated precoding design for downlink wireless network
virtualization (WNV) in  a multi-cell multiple-input multiple-output (MIMO)
network with  imperfect channel state information (CSI). In our WNV framework,
an infrastructure provider (InP) owns each base station that is shared by
several service providers (SPs) oblivious of each other. The SPs design their
precoders as virtualization demands for user services, while the InP designs
the actual precoding solution to meet the service demands from the SPs. Our
aim is to minimize the long-term time-averaged expected precoding deviation
over MIMO fading channels, subject to both per-cell long-term and short-term
transmit power limits. We propose an online coordinated precoding algorithm
for virtualization, which provides a fully distributed semi-closed-form precoding
solution at each cell, based only on the current imperfect CSI without any
CSI exchange across cells. Taking into account the two-fold impact of imperfect
CSI on both the InP and the SPs, we show that our proposed algorithm is within
an $O(\delta)$ gap from the optimum over any time horizon, where $\delta$
is a CSI inaccuracy indicator. Simulation results validate the performance
of our proposed algorithm under two commonly used precoding techniques in
a typical urban micro-cell network environment.\end{abstract}

\section{Introduction}
Wireless network virtualization (WNV) aims at sharing common network infrastructure
among multiple virtual networks to reduce the capital and operational expenses
of wireless networks~\cite{WNVSvy}. In WNV, the infrastructure provider (InP)
virtualizes the physical wireless infrastructure and radio resource into
virtual slices, while the service providers (SPs) lease these virtual slices
and serve their subscribing users under their respective management and requirements
\cite{WLVWNV}. Different from wired network virtualization, WNV concerns
the sharing of both the wireless hardware and the radio spectrum. The random
nature of the wireless medium brings new challenges to guarantee the isolation
of virtual networks~\cite{sliceWNV}.

In this work, we focus on downlink WNV in a multi-cell multiple-input multiple-output
(MIMO) system, where multiple InP-owned base stations (BSs), each with multiple
antennas, are shared by multiple SPs to serve their subscribing users. Most
prior studies on MIMO WNV considered \textit{strict} physical isolation,
where the InP allocates exclusive subsets of antennas or orthogonal sub-channels
to each SP \cite{rp}-\nocite{EE17}\nocite{V5G}\nocite{CRAN}\nocite{WNVNOMA}\cite{Sgame}.
This physical isolation approach is inherited from wired network virtualization
\cite{WNVart}. It does not take full advantage of spatial spectrum sharing
enabled by MIMO precoding. In contrast, in \cite{MPaper}, a spatial isolation
approach was proposed for a single-cell MIMO system, where the SPs share
all antennas and spectrum resource simultaneously. The SPs design their respective
virtual precoding matrices as virtualization demands, based on their users'
local channel states and service needs. Since the SPs are oblivious of each
other, direct implementation of their requested precoding matrices would
induce an unacceptable amount of interference to each other. Instead, the
InP designs the actual downlink precoding to mitigate the inter-SP interference
while satisfying the SPs' virtualization demands. It has been demonstrated
in \cite{MPaper} that, with an optimally designed InP precoding matrix, such
a \textit{spatial} isolation approach substantially outperforms the physical
isolation approach. Adopting the same virtualization approach, in this work,
we consider WNV in a multi-cell MIMO system.

All of the above works on MIMO WNV have focused on per-slot design optimization
problems, subject to a per-slot transmit power constraint. Besides this short-term
power limit, the long-term average transmit power is an important indicator
of energy usage \cite{EE}. Under the long-term power limit, the virtualization
design becomes a stochastic optimization problem, depending on the underlying
channel state variation over time. In this work, we consider the \textit{online}
optimization of MIMO WNV under both short-term and long-term power constraints.
Our objective is to design optimal global downlink precoding at the InP to
serve all users simultaneously, given the set of local virtualization demands
from SPs based on their users' service needs. Note that although the SPs
are oblivious of each other when providing their virtualization demands,
the InP needs to handle both inter-SP and inter-cell interference while trying
to meet each SP's virtualization demand. Thus, the optimization criterion
is the long-term time-averaged deviation between the SPs' virtualization
demands and the actual received signals at their users.

In a traditional non-virtualized multi-cell network, \textit{coordinated}
precoding across the BSs has been widely adopted as a key technique to mitigate
inter-cell interference  \cite{W.Yu-CoOD}\nocite{CShen12}\nocite{X.Wang-CoOD}\nocite{LuoMMSE11}-\cite{T.Queck-CoOD}.
It provides significant performance improvement over non-coordinated networks.
Furthermore, coordinated precoding only requires precoding coordination without
the need to share transmit data across cells or stringent synchronization
among cells. Therefore, in this work, we use coordinated precoding at the
InP to mitigate inter-cell interference. Although offline multi-cell coordinated
precoding has been extensively studied in non-virtualized wireless networks,
new challenges arise for online multi-cell coordinated MIMO WNV. Specifically,
since the SPs are oblivious of each other, it is not effective for each SP
to manage inter-cell interference on its own. Therefore, we consider the
scenario where each SP only has the channel state information (CSI) of its
subscribing users in each cell (\ie those users in a virtual cell), without
access to the CSI of other SPs' users within the cell or users in the other
cells. As a result, their virtual precoding demands sent to the InP do not
consider either inter-SP or inter-cell interference. Thus, the InP must intelligently
design the online precoder to manage the interference among different SPs
and cells while trying to meet the SPs' virtual precoding demands in the
long run. This online virtualized coordinated precoding design problem is
more challenging than the traditional one in the non-virtualized scenario.

Besides the challenges mentioned above, in practical wireless systems, there
are unavoidable CSI errors introduced by channel estimation, quantization,
and imperfect feedback, especially for MIMO fading channels. These errors
may cause significant precoding performance degradation. Thus, it is important
to account for such CSI errors in our online virtualization design and analyze
the impact of CSI errors on the virtualization performance.   Some existing
MIMO WNV solutions can accommodate imperfect CSI \cite{EE17}, \cite{CRAN},
\cite{WNVNOMA}, \cite{MPaper}. However, these works do not allow the SPs
to provide their virtualization demands based on the available CSI adaptively.
Therefore, the impact of imperfect CSI is only on the InP's virtualization
strategy. In contrast, in our problem, imperfect CSI has a two-fold impact
on both the InP and the SPs, as both of them rely on the CSI to design the
actual and virtual precoding matrices. 

In this paper, we present an online design of downlink MIMO WNV over MIMO
fading channels in the presence of imperfect CSI. To facilitate the modeling,
formulation, and analysis, we first focus on the single-cell case and then
extend our study to the multi-case scenario. The main contributions of this
paper are summarized below:

\begin{itemize}

\item We use the spatial isolation approach to formulate the downlink multi-cell
MIMO WNV as an online coordinated precoding problem for efficient spatial
and spectrum resource sharing, subject to both short-term and long-term transmit
power constraints at each cell. Each SP locally designs its precoder in a
cell based on the imperfect local CSI without the knowledge of other SPs'
users in this cell or users in other cells. The InP designs the global precoder
based on the imperfect global CSI. The objective is to minimize the long-term
time-averaged expected deviation between the received signals from the InP's
actual precoder and the SPs' virtualization demands, which implicitly mitigates
both inter-SP and inter-cell interference. 

\item Assuming MIMO fading channel with a bounded CSI error, we propose an
online multi-cell coordinated MIMO WNV algorithm, where we develop new techniques
to extend the standard Lyapunov optimization to handle imperfect CSI. Our
proposed algorithm provides downlink precoding based only on the current
imperfect CSI. Furthermore, our online precoding solution is fully distributed
and in semi-closed form, which can be implemented at each cell without any
CSI exchange across~cells.

\item We provide in-depth performance analysis of our proposed algorithm.
We show that, over any given time horizon, our proposed algorithm using only
the current imperfect CSI can achieve a performance arbitrarily close to
an $\mathcal{O}(\delta)$ gap to the optimal performance under perfect CSI,
where $\delta$ is a normalized measure of CSI error. Our performance analysis
takes into account the effect of imperfect CSI on both the InP and the SPs.
To the best of our knowledge, this is the first work to analyze such two-fold
impact of imperfect CSI on the design performance.

\item Our simulation study under typical urban micro-cell Long-Term Evolution
(LTE) network settings demonstrates that the proposed algorithm has a fast
convergence rate and is robust to imperfect CSI. We further demonstrate the
performance advantage of our proposed spatial virtualization approach over
the traditional physical isolation approach.

\end{itemize}

The rest of the paper is organized as follows. In Section \ref{Sec:Related
Works}, we present the related work. Section \ref{Sec:System Model and Problem
Formulation} describes the single-cell system model and problem formulation.
In Section \ref{Sec:Online MIMO WNV Algorithm}, we present our online  algorithm
and precoding solution for the single-cell case. Performance  bounds are
provided in Section \ref{Sec:Performance Analysis}. In Section~\ref{sec:multi-cell},
we extend the virtualization model and problem to the multi-cell case, present
the proposed online algorithm, and provide performance analysis. Simulation
results are presented in Section \ref{Sec:Simulation Results}, followed by
concluding remarks in Section \ref{Sec:Conclusions}.

\textit{Notations}: The transpose, complex conjugate, Hermitian transpose,
inverse, Euclidean norm, Frobenius norm, trace, and the $(i,j)$ element of
a matrix $\mathbf{A}$ are denoted by $\mathbf{A}^T$, $\mathbf{A}^*$, $\mathbf{A}^{H}$,
$\mathbf{A}^{-1}$, $\Vert\mathbf{A}\Vert_2$, $\Vert \mathbf{A} \Vert_{F}$,
$\text{Tr}\{\mathbf{A}\}$, and $[\mathbf{A}]_{i,j}$, respectively. A positive
definite matrix is denoted as $\mathbf{A}\succ\mathbf{0}$. The notation $\blkdiag\{\mathbf{A}_1,\dots,\mathbf{A}_n\}$
denotes a block diagonal matrix with diagonal elements being matrices $\mathbf{A}_1,\dots\mathbf{A}_n$,
$\mathbf{I}$ denotes an identity matrix, $\mathbb{E}\{\cdot\}$ denotes expectation,
and $\Re\{\cdot\}$ denotes the real part of the enclosed parameter. For $\mathbf{g}$
being an $n\times1$ vector, $\mathbf{g}\sim\mathcal{CN}(\mathbf{0},\sigma^2\mathbf{I})$
means that $\mathbf{g}$ is a circular complex Gaussian random vector with
mean $\mathbf{0}$ and variance $\sigma^2\mathbf{I}$. 

\section{Related Work}
\label{Sec:Related Works}

Among existing works on MIMO WNV that enforce strict physical  isolation,
\cite{rp} and \cite{EE17} studied the problems of throughput maximization
and energy minimization, respectively. Both considered the orthogonal frequency
division multiple access system with massive MIMO. A two-level hierarchical
auction architecture was proposed in \cite{V5G} to allocate exclusive sub-carriers
among the SPs. The uplink resource allocation problems were investigated
in \cite{CRAN} and \cite{WNVNOMA}, combining MIMO WNV with the cloud radio
networks and non-orthogonal multiple access techniques, respectively. Antenna
allocation through pricing was studied in~\cite{Sgame} for virtualized massive
MIMO systems. The spatial isolation approach was first proposed  in \cite{MPaper},
where virtualization is achieved by MIMO precoding design. It has been demonstrated
that this approach substantially outperforms the strict physical isolation
approach. All the above works on MIMO WNV focus on per-slot problems in single-cell
systems.

Various online transmission and resource allocation problems in non-virtualized
wireless systems have been studied in \cite{Dong}\nocite{Two-Hop}\nocite{HY}\nocite{green}-\cite{UW}.
The general Lyapunov optimization technique \cite{Neely} was applied to develop
the online schemes in \cite{Dong}\nocite{Two-Hop}-\cite{HY}. Online power
control for wireless transmission with energy harvesting and storage was
studied for point-to-point transmission \cite{Dong} and two-hop relaying
\cite{Two-Hop}.  Dynamic precoding design  for point-to-point MIMO systems
was studied in \cite{HY}, by extending standard Lyapunov optimization to
deal with imperfect CSI. Online convex optimization technique \cite{OCO}
was applied for MIMO uplink precoding design  in \cite{green} and \cite{UW}.
Recently, the Lyapunov optimization technique and online convex optimization
technique were used to design online downlink precoding for MIMO WNV with
perfect CSI \cite{GLOBECOM19} and delayed CSI \cite{SPAWC20}, respectively.
Neither of their CSI models apply to the present work, and furthermore they
are still limited to single-cell systems.

In this work, we study online coordinated multi-cell WNV over MIMO fading
channels with imperfect CSI. The work in \cite{HY} is the most related to
our problem. However, our MIMO virtualization problem is more challenging
with several key differences: 1) we design MIMO precoding for virtualization,
which features a  virtualization demand and response mechanism between the
InP and the SPs; 2) the SPs are oblivious of each other but share antennas
and spectrum resource provided by the InP; 3) both the InP and the SPs design
either actual precoding or virtual demands based on imperfect CSI; 4) our
online coordinated precoding is for virtualization in multi-cell systems,
where we need to consider inter-cell interference and per-cell transmit power
limit. These unique features for virtualization bring new challenges to the
design of online algorithm and the performance analysis, which were not considered
in \cite{HY}. In particular, imperfect CSI has a two-fold impact on both
the InP and the SPs for their respective precoding designs. New techniques
need to be developed to bound the virtualization performance measured by
the difference between the SPs' virtualization demands and the InP's actual
precoding outcome. Furthermore, the online algorithm and performance analysis
in \cite{HY} are confined to point-to-point MIMO systems, while we consider
a general multi-cell network.

In traditional non-virtualized cellular networks, multi-cell coordinated
precoding has been widely considered to mitigate inter-cell interference
for significant performance improvement \cite{W.Yu-CoOD}\nocite{CShen12}\nocite{X.Wang-CoOD}\nocite{LuoMMSE11}-\cite{T.Queck-CoOD}.
All of the existing works  focus on per-slot coordinated precoding design
problems with given CSI in the current time slot and per-slot maximum transmit
power limit. Coordinated precoding problems were studied for various design
objectives, such as weighted sum transmit power minimization with perfect
CSI \cite{W.Yu-CoOD} and imperfect CSI \cite{CShen12}, weighted sum-rate
maximization \cite{X.Wang-CoOD}, \cite{LuoMMSE11}, and minimum SINR maximization
[18]. To the best of our knowledge, our work is the first to study online
coordinated precoding design for virtualization over fading channels with
both per-slot and long-term transmit power constraints, while accommodating
imperfect CSI. 

\section{System Model and Problem Formulation}
\label{Sec:System Model and Problem Formulation}

\subsection{System Model}

We consider a virtualized MIMO cellular network formed by one InP and $M$
SPs. In each cell, the InP owns the BS and performs virtualization for data
transmission. The SPs are oblivious of each other and serve their subscribing
users. Other functional structures of the network, including the core network
and computational resource, are assumed to be already virtualized.

We first consider the network virtualization design in a single-cell MIMO
system. The virtualization model and problem formulation will be extended
to the multi-cell case in Section~\ref{sec:multi-cell}. Consider downlink
transmissions in a virtualized cell, where the InP-owned BS is equipped with
$N$ antennas. The $M$ SPs share the antennas at the BS and the spectrum resource
provided by the InP. Each SP $m$ serves $K_m$ users. The total number of
users in the cell is $K=\sum_{m\in\mathcal{M}}K_m$. We denote the following
set of indexes $\mathcal{N}=\{1,\dots,N\}$, $\mathcal{M}=\{1,\dots,M\}$,
$\mathcal{K}_m=\{1,\dots,K_m\}$, and $\mathcal{K}=\{1,\dots,K\}$.

We consider a time-slotted system with each time slot indexed by $t\in\{0,1,\dots,T-1\}$.
Let $\mathbf{H}_m(t)\in\mathbb{C}^{K_m\times{N}}$ denote the  channel state
between the BS and $K_m$ users served by SP $m$ at time $t$. Let $\mathbf{H}(t)=[\mathbf{H}_1^H(t),\dots,\mathbf{H}_M^H(t)]^H\in\mathbb{C}^{K\times{N}}$
denote the channel state between the BS and all $K$ users at time $t$. We
assume a block fading channel model, where the sequence of channel state
$\{\mathbf{H}(t)\}$ over time $t$ is independent and identically distributed
(i.i.d.). The distribution of $\mathbf{H}(t)$ can be arbitrary and is unknown
at the BS. We assume that the channel gain is bounded by constant $B\ge0$
at any time $t$, \ie
\begin{align}
        \Vert\mathbf{H}(t)\Vert_F\leq B,\quad \forall t. \label{eq:asm1}
\end{align}

\begin{figure}[!t]
\vspace{0mm}
\centering
\includegraphics[width=1\linewidth,trim=200 130 200 130,clip]{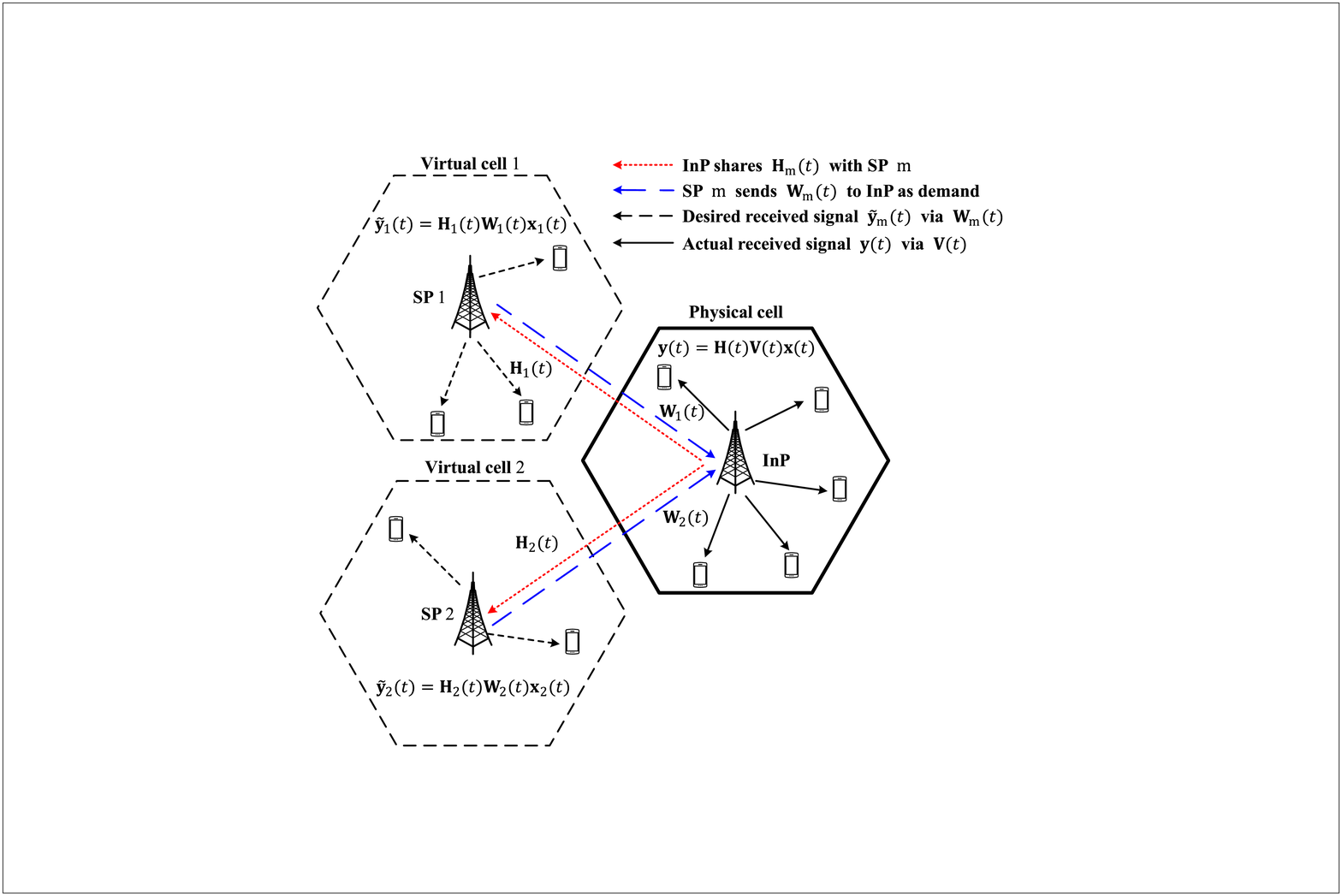}
\vspace{-6mm}
\caption {An illustration of MIMO virtualization in a cell with one InP and
two SPs each serving its users in a virtual cell.}\label{VMIMO}
\vspace{-2mm}
\end{figure}

We adopt the spatial virtualization approach first proposed in \cite{MPaper},
which is illustrated in Fig.~\ref{VMIMO}. In the idealized case when the
perfect CSI is available at each time $t$, the InP shares with SP $m$ the
channel state $\mathbf{H}_m(t)$ between the BS and its $K_m$ users and allocates
transmission power $P_m$ to the SP. Based on $\mathbf{H}_m(t)$, each SP $m$
designs its precoding matrix $\mathbf{W}_m(t)\in\mathbb{C}^{N\times K_m}$,
subject to the transmission power limit $\Vert\mathbf{W}_m(t)\Vert_F^2\leq
P_m$. The design of $\mathbf{W}_m(t)$ is solely based on the service needs
of SP $m$'s users, without considering the existence of the other SPs sharing
the same BS antennas and spectrum resource. Each SP $m$ then sends $\mathbf{W}_m(t)$
as its virtual precoding demand to the InP. For SP $m$, the \textit{desired}
received signal vector (noiseless) $\tilde{\mathbf{y}}_m$ (at $K_m$ users)
is given by 
\begin{align*}
        \tilde{\mathbf{y}}_m(t)=\mathbf{H}_m(t)\mathbf{W}_m(t)\mathbf{x}_m(t)
\end{align*}
where $\mathbf{x}_m(t)$ is the symbol vector to $K_m$ users. Define the desired
received signal vector at all $K$ users in the network as $\tilde{\mathbf{y}}(t)\triangleq[\tilde{\mathbf{y}}_1^H(t),\dots,\tilde{\mathbf{y}}_M^H(t)]^H$,
we have $\tilde{\mathbf{y}}(t)= \mathbf{D}(t)\mathbf{x}(t)$, where $\mathbf{D}(t)\triangleq\blkdiag\{\mathbf{H}_1(t)\mathbf{W}_1(t),\dots,\mathbf{H}_M(t)\mathbf{W}_M(t)\}\in\mathbb{C}^{K\times{K}}$
is the virtualization demand from all SPs, and  $\mathbf{x}(t)\triangleq[\mathbf{x}_1^H(t),\dots,\mathbf{x}_M^H(t)]^H$
contains the symbols to all $K$ users, which are assumed to have unit power
and be independent to each other, \ie $\mathbb{E}\{\mathbf{x}(t)\mathbf{x}^H(t)\}=\mathbf{I},\forall{t}$.

At each time $t$, the InP designs the actual  precoding matrix $\mathbf{V}(t)\triangleq[\mathbf{V}_1(t),\dots,\mathbf{V}_M(t)]\in\mathbb{C}^{N\times{K}}$,
where $\mathbf{V}_m(t)\in\mathbb{C}^{N\times K_m}$ is the actual precoding
matrix for SP $m$. The \textit{actual} received signal vector (noiseless)
$\mathbf{y}_m(t)$ at SP $m$'s $K_m$ users is given by
\begin{align*}
        \mathbf{y}_m(t)=\mathbf{H}_m(t)\mathbf{V}_m(t)\mathbf{x}_m(t)+\sum_{i\in\mathcal{M},i\neq{m}}\mathbf{H}_m\mathbf{V}_i(t)\mathbf{x}_i(t),
\end{align*}
where the second term is the inter-SP interference from the other SPs to
SP $m$'s users. The actual received signal vector at all $K$ users is given
by $\mathbf{y}(t)=[\mathbf{y}_1^H(t),\dots,\mathbf{y}_M^H(t)]^H=\mathbf{H}(t)\mathbf{V}(t)\mathbf{x}(t)$.

\subsection{Problem Formulation}

For downlink MIMO WNV, the InP designs precoding matrix $\mathbf{V}(t)$ to
perform MIMO virtualization. Note that while each SP $m$ designs its virtual
precoding matrix $\mathbf{W}_m(t)$ without considering the inter-SP interference,
the InP designs the actual precoding matrix $\mathbf{V}(t)$ to mitigate the
inter-SP interference, in order to meet the virtualization demand $\mathbf{D}(t)$
of all SPs.

With the InP's actual precoding matrix $\mathbf{V}(t)$ and each SP $m$'s
virtual precoding matrix $\mathbf{W}_m(t)$, the expected deviation of the
actual received signal vector at all $K$ users from the desired one is $\mathbb{E}\{
\Vert\mathbf{y}(t)-\tilde{\mathbf{y}}(t)\Vert_2^2\}=\mathbb{E}\{\Vert\mathbf{H}(t)\mathbf{V}(t)-\mathbf{D}(t)\Vert_F^2\}$.

The goal at the InP is to optimize  MIMO precoding to minimize the long-term
time-averaged expected precoding deviation from the virtualization demands,
subject to both long-term and short-term transmit power constraints. The
optimization problem is formulated as follows:
\begin{align}
        \textbf{P1}:\quad\min_{\{\mathbf{V}(t)\}}\quad&\lim_{T\to\infty}\frac{1}{T}\sum_{t=0}^{T-1}\mathbb{E}\{\Vert\mathbf{H}(t)\mathbf{V}(t)-\mathbf{D}(t)\Vert_F^2\}\notag\\
        \text{s.t.} ~\quad &\lim_{T\to\infty}\frac{1}{T}\sum_{t=0}^{T-1}\mathbb{E}\{\Vert\mathbf{V}(t)\Vert_F^2\}\leq\bar{P},\label{VM2}\\
        &\Vert\mathbf{V}(t)\Vert_F^2 \leq P_{\text{max}}\label{VM3}
\end{align}
where $\bar{P}$ is the long-term average transmit power limit, and $P_{\text{max}}$
is the per-slot maximum transmit power limit at the BS. Both power limits
are set by the InP, and we assume $\bar{P}\leq P_{\text{max}}$.

Since channel state $\mathbf{H}(t)$ is random, \textbf{P1} is a stochastic
optimization problem. The problem is challenging to solve, especially when
the distribution of $\mathbf{H}(t)$ is unknown, especially in massive MIMO
systems with a large number of antennas and users. In addition, the instantaneous
channel state cannot be obtained accurately in practical systems. Typically,
the InP only has an inaccurate estimate of the channel state $\hat{\mathbf{H}}(t)$
at each time  $t$. With a given channel estimation quality, we assume the
normalized CSI inaccuracy is bounded by a constant $\delta\geq0$ at any time
$t$, given by
\begin{align}
        \frac{\Vert\tilde{\mathbf{H}}_m(t)\Vert_F}{\Vert\mathbf{H}_m(t)\Vert_F}\leq\delta,\quad\forall{m}\in\mathcal{M},\quad\forall{t}\label{eq:asm2}
\end{align}
where $\tilde{\mathbf{H}}_m(t)\triangleq\mathbf{H}_m(t) - \hat{\mathbf{H}}_m(t)$
is the channel estimation error, with $\hat{\mathbf{H}}_m(t)$ being the estimated
channel state of SP~$m$'s users. By (\ref{eq:asm1}) and (\ref{eq:asm2}),
the estimated channel gain is bounded~by
\begin{align}
        \Vert\hat{\mathbf{H}}(t)\Vert_F\leq\Vert\mathbf{H}(t)\Vert_F+\Vert\tilde{\mathbf{H}}(t)\Vert_F\leq{B}(1+\delta),\quad\forall{t}.\label{Hhat}
\end{align}

Using the estimated channel state $\hat{\mathbf{H}}_m(t)$ shared by the InP
at time $t$, each SP $m$ designs its virtual precoding matrix, denoted by
$\hat{\mathbf{W}}_m(t)$. As a result, the InP receives  virtualization demand
$\hat{\mathbf{D}}(t)\triangleq\blkdiag\{\hat{\mathbf{H}}_1(t)\hat{\mathbf{W}}_1(t),\dots,\hat{\mathbf{H}}_M(t)\hat{\mathbf{W}}_M(t)\}$
from the SPs based on the imperfect CSI. Using $\hat{\mathbf{H}}(t)$ and
$\hat{\mathbf{D}}(t)$, the InP then designs the actual precoding matrix,
denoted by $\hat{\mathbf{V}}(t)$.

Our goal is to develop an online MIMO WNV algorithm  based on $\hat{\mathbf{H}}(t)$
and $\hat{\mathbf{D}}(t)$ to find a precoding solution $\{\hat{\mathbf{V}}(t)\}$
to \textbf{P1} under the unknown channel distribution $\mathbf{H}(t)$.

\section{Online Single-Cell MIMO WNV Algorithm}
\label{Sec:Online MIMO WNV Algorithm}

In this section, we present a new online precoding algorithm for MIMO WNV
that is developed based on the Lyapunov optimization technique. Note that
the standard Lyapunov optimization relies on accurate system state \cite{Neely},
which is not applicable to our problem. Instead, we develop new techniques
to accommodate imperfect CSI in designing the online algorithm at both the
InP and the SPs.

\subsection{Online Optimization Formulation}
\label{sec:Online Optimization Formulation}

To design an online algorithm for solving \textbf{P1}, we introduce a virtual
queue $Z(t)$ for the long-term average transmit power constraint (\ref{VM2})
with the updating rule given by
\begin{align}
        Z(t+1) = \max\{Z(t)+ \Vert \hat{\mathbf{V}}(t) \Vert_F^2-\bar{P},0\}.\label{eq:VQ}
\end{align}
Define $L(t)\triangleq\frac{1}{2} Z^2(t)$ as the quadratic Lyapunov function
and $\Delta(t)\triangleq L(t+1)-L(t)$ as the corresponding one-slot Lyapunov
drift at time $t$. By the theory of Lyapunov optimization, \textbf{P1} can
be converted to  minimizing the objective function while stabilizing the
virtual queue in (\ref{eq:VQ}), which can be further converted to  minimizing
a drift-plus-penalty (DPP) metric \cite{Neely}. The DPP metric is defined
as $\mathbb{E}\{\Delta(t)|Z(t)\}+U\mathbb{E}\{\hat{\rho}(t)|Z(t)\}$,
where $\hat{\rho}(t)\triangleq\Vert\hat{\mathbf{H}}(t)\hat{\mathbf{V}}(t)-\hat{\mathbf{D}}(t)\Vert_F^2$
is the penalty cost representing the precoding deviation from the virtualization
demands based on imperfect CSI, and $U>0$ is the relative weight. The DPP
metric is a weighted sum of the expected Lyapunov drift $\Delta(t)$ and the
precoding deviation $\hat{\rho}(t)$ under the current estimated channel state
$\hat{\mathbf{H}}(t)$, conditioned on the current virtual queue $Z(t)$. Minimizing
the DPP metric directly is still difficult due to the dynamics involved in
the Lyapunov drift $\Delta(t)$. Instead, we first provide an upper bound
for the DPP metric in the following lemma. 

\begin{lem}\label{lmDPPUB}
At each time $t$, for any precoding design of $\hat{\mathbf{V}}(t)$, the
DPP metric has the following upper bound for all $Z(t)$ and $U> 0$
\begin{align}
        &\mathbb{E}\{\Delta(t)|Z(t)\}+U\mathbb{E}\{\hat{\rho}(t)|Z(t)\}\notag\\
        &\le{S}+U\mathbb{E}\left\{\hat{\rho}(t)|Z(t)\right\}+Z(t)\mathbb{E}\{\Vert\hat{\mathbf{V}}(t)\Vert_F^2-\bar{P}|Z(t)\}\label{DPPUB}
\end{align}
where $S\triangleq\frac{1}{2}\max\left\{(P_{\text{max}}-\bar{P})^2,\bar{P}^2\right\}$.
\end{lem}
\textit{Proof:} From the virtual queue dynamics in (\ref{eq:VQ}), we have
\begin{align}
        \Delta(t)&=\frac{1}{2}\left((\max\{{Z}(t)+\Vert\hat{\mathbf{V}}(t)\Vert_F^2-\bar{P},0\})^2-Z^2(t)\right)\notag\\
        &\leq \frac{1}{2}\left( ( Z(t)+\Vert\hat{\mathbf{V}}(t)\Vert_F^2-\bar{P})^2-Z^2(t)\right)\notag\\
        &=\frac{1}{2}(\Vert\hat{\mathbf{V}}(t)\Vert_F^2-\bar{P})^2+Z(t)(\Vert\hat{\mathbf{V}}(t)\Vert_F^2-\bar{P}).\label{eq:lm1-1}
\end{align}
By the short-term transmit power constraint in (\ref{VM3}), we have 
\begin{align}
        (\Vert\hat{\mathbf{V}}(t)\Vert_F^2-\bar{P})^2\leq\max\{(P_{\text{max}}-\bar{P})^2,\bar{P}^2\}.\label{eq:lm1-2}
\end{align}
Taking the conditional expectation at both sides of (\ref{eq:lm1-1}) for
given $Z(t)$ and considering (\ref{eq:lm1-2}), we have the upper bound of
the per-slot conditional expected Lyapunov drift $\Delta(t)$ given by
\begin{align}
        &\mathbb{E}\{\Delta(t) | Z(t)\} \leq S +\mathbb{E}\left\{Z(t)(\Vert\hat{\mathbf{V}}(t)\Vert_F^2-\bar{P})|Z(t)\right\}.\label{eq:lm1-3}
\end{align}
Adding $U \mathbb{E}\{\hat{\rho}(t) |Z(t)\}$ on both sides of (\ref{eq:lm1-3}),
we have (\ref{DPPUB}).\endIEEEproof

Using Lemma \ref{lmDPPUB}, instead of directly minimizing the DPP metric,
we minimize its upper bound in (\ref{DPPUB}), which is no longer a function
of $\Delta(t)$. Specifically, given $\hat{\mathbf{H}}(t)$ at time $t$, we
consider the per-slot version of the upper bound in (\ref{DPPUB}) as the
optimization objective by removing the conditional expectation. By removing
the constant terms, the resulting per-slot optimization  problem is given~by
\begin{align}
        \textbf{P2}:~\min_{\hat{\mathbf{V}}(t)} ~ & U\Vert \hat{\mathbf{H}}(t)\hat{\mathbf{V}}(t)-\hat{\mathbf{D}}(t)\Vert_F^2+Z(t)\Vert\hat{\mathbf{V}}(t)\Vert_F^2\notag\\
        \qquad\text{s.t.}~~&\Vert\hat{\mathbf{V}}(t)\Vert_F^2 \leq P_{\text{max}}.\label{sl2}
\end{align}
Note that \textbf{P2} is a per-slot precoding optimization problem under
the current estimated channel state $\hat{\mathbf{H}}(t)$ and the virtual
queue $Z(t)$, subject to the per-slot maximum transmit power constraint (\ref{sl2}).
Compared with the original \textbf{P1}, we change the long-term time-averaged
expected objective to the per-slot version of DPP metric in \textbf{P2},
where the long-term average transmit power constraint (\ref{VM2}) is converted
into maintaining the queue stability in $Z(t)$ as part of the DPP metric.
Next, we solve \textbf{P2} to obtain the optimal precoding matrix $\hat{\mathbf{V}}^{\star}(t)$
for \textbf{P2} at each time $t$, and then update the virtual queue $Z(t)$
according to its queue dynamics in (\ref{eq:VQ}). An outline of the proposed
online algorithm is given in Algorithm \ref{alg:1}.

\setlength{\textfloatsep}{0.5cm}
\begin{algorithm}[t]
\begin{small}
\caption{Outline of Online MIMO WNV Algorithm }\label{alg:1}
\begin{algorithmic}[1]
\STATE Set $U>0$ and $Z(0)=0$.
\STATE At each time $t$, obtain $\hat{\mathbf{H}}(t)$ and $Z(t)$.
\STATE Solve  \textbf{P2} for $\hat{\mathbf{V}}^{\star}(t)$ (see Section
\ref{Subsec:precoding}).
\STATE Update  $Z(t+1) = \max\{Z(t)+ \Vert \hat{\mathbf{V}}^{\star}(t)\Vert_F^2-\bar{P},0\}$.
\end{algorithmic} 
\end{small}
\end{algorithm}

\subsection{Online Precoding Solution to \textbf{P2}}\label{Subsec:precoding}

Now we present a semi-closed-form solution to \textbf{P2}. Without causing
any ambiguity, for notation simplicity, we omit the time index $t$ in solving
\textbf{P2}. Note that \textbf{P2} is essentially a constrained regularized
least square problem. Since \textbf{P2} is a convex optimization problem
satisfying Slater's condition, the strong duality holds. We solve \textbf{P2}
using the Karush-Kuhn-Tucker (KKT) conditions \cite{Boyd}. The Lagrangian
for \textbf{P2} is given by
\begin{align*}
        &L(\hat{\mathbf{V}},\lambda)=U\Vert\hat{\mathbf{H}}\hat{\mathbf{V}}-\hat{\mathbf{D}}\Vert_F^2+Z\Vert\hat{\mathbf{V}}\Vert_F^2+\lambda(\Vert\hat{\mathbf{V}}\Vert_F^2-P_{\text{max}})\\
        &=U(\tr\{\hat{\mathbf{H}}^H\hat{\mathbf{H}}\hat{\mathbf{V}}\hat{\mathbf{V}}^H\}+\tr\{\hat{\mathbf{D}}\hat{\mathbf{D}}^H\}-\tr\{\hat{\mathbf{H}}^H\hat{\mathbf{D}}\hat{\mathbf{V}}^H\}\notag\\
        &\quad-\tr\{\hat{\mathbf{H}}\hat{\mathbf{V}}\hat{\mathbf{D}}^H\})+(Z+\lambda)\tr\{\hat{\mathbf{V}}\hat{\mathbf{V}}^H\}-\lambda{P}_{\text{max}}
\end{align*}
where $\lambda$ is the Lagrange multiplier associated with the maximum power
constraint (\ref{sl2}). The gradient of $L(\hat{\mathbf{V}},\lambda)$ w.r.t.
$\mathbf{\hat{V}}^*$ is given by 
\begin{align}
       \nabla_{\mathbf{\hat{V}}^*} L(\hat{\mathbf{V}},\lambda) = U(\hat{\mathbf{H}}^H\hat{\mathbf{H}}\hat{\mathbf{V}}-\hat{\mathbf{H}}^H\hat{\mathbf{D}})+(Z+\lambda)\hat{\mathbf{V}}\label{GD}
\end{align}
where the following fact is used: $\nabla_{\mathbf{B}^*}\tr\{\mathbf{AB}^H\}=\mathbf{A}$
and $\nabla_{\mathbf{B}^*}\tr\{\mathbf{AB}\}=\mathbf{0}$ \cite{CGD}. The
optimal solution to $\textbf{P2}$ can be obtained by solving the KKT conditions,
given by
\begin{align}
        \left(\hat{\mathbf{H}}^H\hat{\mathbf{H}} + \frac{Z+\lambda^{\star}}{U}\mathbf{I}\right)\hat{\mathbf{V}}^{\star}&=\hat{\mathbf{H}}^H\hat{\mathbf{D}},\label{kkt1}\\
        \Vert\hat{\mathbf{V}}^{\star}\Vert_F^2 - P_{\text{max}}&\leq 0,\label{kkt2}\\
        \lambda^{\star}&\geq{0},\label{kkt3}\\
        \lambda^{\star}( \Vert\hat{\mathbf{V}}^{\star}\Vert_F^2 - P_{\text{max}})&=0\label{slack}
\end{align}
 where (\ref{kkt1}) is obtained by setting $\nabla_{\mathbf{\hat{V}}^*} L(\hat{\mathbf{V}},\lambda)$
in (\ref{GD}) to $\mathbf{0}$. Note that, by (\ref{eq:VQ}), the virtual queue
is nonnegative, \ie $Z\geq0$. Thus, we derive the optimal solution in the
following two cases.

\subsubsection{$Z+\lambda^{\star}>0$}

From (\ref{kkt1}) and $\hat{\mathbf{H}}^H\hat{\mathbf{H}}+\frac{Z+\lambda^{\star}}{U}\mathbf{I}\succ\mathbf{0}$,
we have
\begin{align}
        \hat{\mathbf{V}}^{\star} =\left( \hat{\mathbf{H}}^H\hat{\mathbf{H}}+\frac{Z+\lambda^{\star}}{U}\mathbf{I}\right)^{-1}\hat{\mathbf{H}}^H\hat{\mathbf{D}}.\label{Vstar}
\end{align}
Depending on $Z$, we determine $\lambda^\star$ in (\ref{Vstar}) in two subcases:
1.\romannum{1}) If $Z>0$: By (\ref{kkt2}) and
(\ref{slack}), we conclude that if $\Vert( \hat{\mathbf{H}}^H\hat{\mathbf{H}}+
\frac{Z}{U}\mathbf{I})^{-1}\hat{\mathbf{H}}^H\hat{\mathbf{D}}\Vert_F^2\leq{P}_{\text{max}}$,
then $\lambda^{\star}=0$; Otherwise, we have $\lambda^{\star}>0$ such that
$\Vert(\hat{\mathbf{H}}^H\hat{\mathbf{H}} + \frac{Z+\lambda^{\star}}{U}\mathbf{I})^{-1}\hat{\mathbf{H}}^H\hat{\mathbf{D}}\Vert_F^2=P_{\text{max}}$.
1.\romannum{2}) If $Z=0$: In this case, $\lambda^{\star}>0$. By (\ref{slack}),
the value of $\lambda^{\star}$ satisfies $\Vert(\hat{\mathbf{H}}^H\hat{\mathbf{H}}+\frac{\lambda^{\star}}{U}\mathbf{I})^{-1}\hat{\mathbf{H}}^H\hat{\mathbf{D}}\Vert_F^2=P_{\text{max}}$.

\subsubsection{$Z=\lambda^{\star}=0$} 

From (\ref{kkt1}), the optimal solution must satisfy
\begin{align}
        \hat{\mathbf{H}}^H\hat{\mathbf{H}}\hat{\mathbf{V}}^{\star}=\hat{\mathbf{H}}^H\hat{\mathbf{D}}.\label{HVD}
\end{align} 
We analyze (\ref{HVD}) in two subcases: 2.\romannum{1}) If $K<N$: $\hat{\mathbf{H}}^H\hat{\mathbf{H}}\in\mathbb{C}^{N\times{N}}$
is a rank-deficient matrix, and thus there are  infinitely many solutions
to $\hat{\mathbf{V}}^{\star}$. We choose $\hat{\mathbf{V}}^{\star}$ that
minimizes $\Vert \hat{\mathbf{V}}^{\star}\Vert_F^2$ subject to (\ref{HVD}).
This problem is an under-determined least square problem with  a closed-form
solution:
\begin{align}
        \hat{\mathbf{V}}^{\star}=\hat{\mathbf{H}}^H\left(\hat{\mathbf{H}}\hat{\mathbf{H}}^H\right)^{-1}\hat{\mathbf{D}}.\label{sol1}
\end{align}
Substitute the above expression of $\hat{\mathbf{V}}^\star$ into the power
constraint in (\ref{kkt2}). If $\Vert\hat{\mathbf{H}}^H(\hat{\mathbf{H}}\hat{\mathbf{H}}^H)^{-1}\hat{\mathbf{D}}\Vert_F^2\leq{P}_{\text{max}}$,
then $\hat{\mathbf{V}}^{\star}$ in (\ref{sol1}) is the optimal solution.
Otherwise, see the discussion in the next paragraph. 2.\romannum{2}) If $K
\ge N$: $\hat{\mathbf{H}}^H\hat{\mathbf{H}}\!\in\!\mathbb{C}^{N\times{N}}$
is full rank\footnote{Since the channels from BS to users are assumed independent,
$\mathbf{H}(t)\in\mathbb{C}^{K\times N}$ is of full rank at each time $t$.
The independent channel assumption is typically satisfied in practice for
users at different locations.}, and we have a unique solution:
\begin{align}
        \hat{\mathbf{V}}^{\star}=\left(\hat{\mathbf{H}}^H\hat{\mathbf{H}}\right)^{-1}\hat{\mathbf{H}}^H\hat{\mathbf{D}}.\label{sol2}
\end{align}
Again, substituting $\hat{\mathbf{V}}^{\star}$ in (\ref{sol2}) into (\ref{kkt2}),
if $\Vert(\hat{\mathbf{H}}^H\hat{\mathbf{H}})^{-1}\hat{\mathbf{H}}^H\hat{\mathbf{D}}\Vert_F^2\leq{P}_{\text{max}}$,
then $\hat{\mathbf{V}}^{\star}$ in (\ref{sol2}) is the optimal solution.

Note that, for both subcases 2.\romannum{1}) and 2.\romannum{2}), if $\hat{\mathbf{V}}^{\star}$
in (\ref{sol1}) or (\ref{sol2}) cannot satisfy the power constraint in (\ref{kkt2}),
it means that the condition in Case 2) does not hold at optimality, and we
have $\lambda^{\star}>0$, \ie the optimal solution is given by (\ref{Vstar}).
 
From the above discussion, if $\lambda^{\star}=0$ at the optimality, we have
a closed-form  solution for $\hat{\mathbf{V}}^{\star}$ in (\ref{sol1}) or
(\ref{sol2}). Otherwise, we have a semi-closed-form solution for $\hat{\mathbf{V}}^{\star}$
in (\ref{Vstar}), where $\lambda^{\star}>0$ can be  obtained through the
bi-section search to ensure the transmit power meets $P_{\text{max}}$ in
(\ref{kkt2}). The computational complexity for calculating $\hat{\mathbf{V}}^{\star}$
is dominated by the matrix inversion, and thus is in the order of $\mathcal{O}(\min(N,K)^3)$.

\section{Performance Bounds for Single-Cell Case}\label{Sec:Performance Analysis}

Different from existing MIMO precoding designs for non-virtualized networks
such as in [27], for the MIMO WNV design, the impact of imperfect CSI on
the system is two-fold at both the InP and the SPs. This brings some unique
challenges in analyzing the performance of the proposed online algorithm.
In this section, we develop new techniques to derive the performance bounds
for our online algorithm.

First, we show in the following lemma that by Algorithm \ref{alg:1}, the
virtual queue $Z(t)$ is upper bounded at each time $t$.

\begin{lem}\label{lmZ}
By Algorithm \ref{alg:1}, $Z(t)$ satisfies
\begin{align}
        Z(t)\leq U  B^2(1+\delta)^2\xi+P_{\text{max}}-\bar{P},\quad\forall{t}\label{BDVQ}
\end{align}
where $\xi\triangleq\sqrt{\frac{N}{\bar{P}}\sum_{m\in\mathcal{M}}P_{m}}$.
\end{lem}
\textit{Proof:} We first omit time index $t$ for simplicity. Let $\hat{\mathbf{H}}^H\hat{\mathbf{H}}=\hat{\mathbf{U}}\hat{\boldsymbol{\Sigma}}\hat{\mathbf{U}}^H$,
where $\hat{\mathbf{U}}$ is an unitary matrix, and $\hat{\boldsymbol{\Sigma}}=\diag(\hat{\sigma}_1,\dots,\hat{\sigma}_{N})$.
It follows that $\hat{\mathbf{H}}^H\hat{\mathbf{H}}+\frac{Z+\lambda^{\star}}{U}\mathbf{I}=\hat{\mathbf{U}}\hat{\boldsymbol{\Phi}}\hat{\mathbf{U}}^H$,
where $\hat{\boldsymbol{\Phi}}=\diag(\hat{\phi}_1,\dots,\hat{\phi}_N)$ and
$\hat{\phi}_n=\hat{\sigma}_n+\frac{Z+\lambda^{\star}}{U},\forall{n}\in\mathcal{N}$.
If $Z>0$, $\hat{\mathbf{V}}^{\star}$ is given in (\ref{Vstar}), and we have
\begin{align}
        \Vert\hat{\mathbf{V}}^{\star}\Vert_F&\leq\left\Vert\hat{\mathbf{U}}\hat{\boldsymbol{\Phi}}^{-1}\hat{\mathbf{U}}^H\right\Vert_F\Vert\hat{\mathbf{H}}\Vert_F\Vert\hat{\mathbf{D}}\Vert_F\stackrel{(a)}{\leq}\frac{U}{Z}\sqrt{N}\Vert\hat{\mathbf{H}}\Vert_F\Vert\hat{\mathbf{D}}\Vert_F\notag\\
        &\stackrel{(b)}{\leq}\frac{U}{Z}B^2(1+\delta)^2\sqrt{N\sum_{m\in\mathcal{M}}P_m}.\!\!\label{eq:lm2-2}
\end{align}
Inequality $(a)$ follows from $\Vert\hat{\mathbf{U}}\hat{\boldsymbol{\Phi}}^{-1}\hat{\mathbf{U}}^H\Vert_F^2=\tr\{\hat{\boldsymbol{\Phi}}^{-2}\}=\sum_{n\in\mathcal{N}}\hat{\phi}_{n}^{-2}$.
Since $\hat{\sigma}_n\geq0,\forall{n}\in\mathcal{N}$, $\lambda^{\star}\geq0$,
and $Z>0$, it follows that $\hat{\phi}_{n}^{-2}\le\frac{U^2}{Z^2},\forall{n}\in\mathcal{N}$,
and therefore $\Vert\hat{\mathbf{U}}\hat{\boldsymbol{\Phi}}^{-1}\hat{\mathbf{U}}^H\Vert_F^2\le{N}\frac{U^2}{Z^2}$.
Inequality $(b)$ follows from (\ref{Hhat}) and by the definition of~$\hat{\mathbf{D}}$
\begin{align}
        \!\!\Vert\hat{\mathbf{D}}\Vert_F^2\leq\!\!\sum_{m\in\mathcal{M}}\Vert\hat{\mathbf{H}}_m\Vert_F^2\Vert\hat{\mathbf{W}}_m\Vert_F^{2}\le{B^2}(1+\delta)^2\sum_{m\in\mathcal{M}}P_m.\!\!\label{lm3-1}
\end{align}

From (\ref{eq:lm2-2}), a sufficient condition to ensure $\Vert \hat{\mathbf{V}}^{\star}\Vert_F^2\leq\bar{P}$
is that the RHS of (\ref{eq:lm2-2}) is less than $\sqrt{\bar{P}}$, which
means $Z\geq UB^2(1+\delta)^2\sqrt{\frac{N}{\bar{P}}\sum_{m\in\mathcal{M}}P_m}$.
Consider time index $t$. If the condition holds, $\Vert \hat{\mathbf{V}}^{\star}
(t)\Vert_F^2\leq\bar{P}$, the virtual queue in (\ref{eq:VQ}) decreases, \ie
$Z(t+1)\leq Z(t)$. Otherwise, the maximum increment of the virtual queue
is $P_{\text{max}}-\bar{P}$, \ie $Z(t+1)\leq Z(t)+P_{\text{max}}-\bar{P}$.
Thus, we have the upper bound of $Z(t)$ in (\ref{BDVQ}) at any time $t\geq{0}$.
\endIEEEproof

Note that Algorithm \ref{alg:1} and the upper bound on the virtual queue
in Lemma \ref{lmZ} are applicable to any precoding schemes adopted by the
SPs. In the following, we consider two common precoding schemes: maximum
ratio transmission (MRT) and zero forcing (ZF). We assume $M_{\text{\begin{tiny}MRT\end{tiny}}}$
SPs adopt MRT precoding and the rest of SPs adopt ZF precoding. We point
out that although the following analysis focuses on the two precoding schemes,
the similar analysis can be extended to other precoding schemes as well.

Let $\mathcal{M}_{\text{\begin{tiny}MRT\end{tiny}}}=\{1,\dots,M_\text{\begin{tiny}MRT\end{tiny}}\}$
be the set of SPs that adopt MRT precoding, with the MRT precoding matrix
given by
\begin{align}
        \hat{\mathbf{W}}_{m}^{\text{\begin{tiny}MRT\end{tiny}}}(t)=\sqrt{P_{m}}\frac{\hat{\mathbf{H}}_{m}^H(t)}{\Vert\hat{\mathbf{H}}_{m}(t)\Vert_F}.\label{MRT}
\end{align}
Each SP $m\in\mathcal{M}\backslash\mathcal{M}_\text{\begin{tiny}MRT\end{tiny}}$
adopts ZF precoding to null the intra-SP interference. We assume $K_m\le
N$ to use ZF precoding. The ZF precoding matrix is given by
\begin{align}
        \hat{\mathbf{W}}_m^{\text{\begin{tiny}ZF\end{tiny}}}(t) = \sqrt{P_{m}}\frac{\hat{\mathbf{H}}_m^H(t)[\hat{\mathbf{H}}_m(t)\hat{\mathbf{H}}_m^H(t)]^{-1}}{\sqrt{\tr\{[\hat{\mathbf{H}}_m(t)\hat{\mathbf{H}}_m^H(t)]^{-1}\}}}.\label{ZF}
\end{align}

With the two precoding matrices in (\ref{MRT}) and (\ref{ZF}), we first quantify
the impact of inaccurate CSI on the SPs' virtualization demands by providing
an upper bound on the deviation between the accurate and inaccurate virtualization
demands $\Vert\mathbf{D}(t)-\hat{\mathbf{D}}(t)\Vert_{F}$, for given CSI
inaccuracy $\delta$ in (\ref{eq:asm2}). This effect of $\delta$ on the precoding
performance is unique to the MIMO WNV system and has not been studied before.

Let $\omega_{m,1}(t),\dots,\omega_{m,K_m}(t)$ be the eigenvalues of $\hat{\mathbf{H}}_m\hat{\mathbf{H}}_m^H$,
and similarly, $\hat{\omega}_{m,1}(t),\dots,\hat{\omega}_{m,K_m}(t)$ the
eigenvalues of $\hat{\mathbf{H}}_m(t)\hat{\mathbf{H}}_m^H(t)$. Define $\hat{B}_m^{\text{\begin{tiny}min\end{tiny}}}\triangleq\min\{\Vert\hat{\mathbf{H}}_m(t)\Vert_F:\forall{t}\}$,
$\hat{\omega}_m^{\text{\begin{tiny}min\end{tiny}}}\triangleq\min\{\hat{\omega}_{m,i}(t):\forall{i}\in\mathcal{K}_m,\forall{t}\}$,
and $\omega_m^{\text{\begin{tiny}min\end{tiny}}}\triangleq\min\{\omega_{m,i}(t):\forall{i}\in\mathcal{K}_m,\forall{t}\}$,
which respectively represent the minimum channel gain of $\hat{\mathbf{H}}_m(t)$,
the minimum energy in the eigen-directions of $\hat{\mathbf{H}}_m(t)$, and
that of $\mathbf{H}_m(t)$. We have the following lemma.

\begin{lem}\label{Dbd}
At each time $t$, the following  hold: 
\begin{align}
\Vert\mathbf{D}(t)\Vert_F&\leq  \zeta B, \label{BDD}\\
\Vert\hat{\mathbf{D}}(t)\Vert_F&\leq \zeta B(1+\delta),\label{BDDhat}\\
\Vert \mathbf{D}(t)-\hat{\mathbf{D}}(t) \Vert_F&\leq\eta{B}\delta,\label{BDDDhat}
\end{align}
where

\noindent  $\eta\!\triangleq\!\!\sqrt{\displaystyle\!\sum_{m\in\mathcal{M}_\text{\begin{tiny}MRT\end{tiny}}}\!\!\left(\!1\!+\!\frac{(2\!+\!\delta)B}{\hat{B}_{m}^{\text{\begin{tiny}min\end{tiny}}}}\!\right)^{\!2}\!\!\!P_{m}\!+\!\!\!\!\sum_{m\in\mathcal{M}\backslash\mathcal{M}_\text{\begin{tiny}MRT\end{tiny}}}\!\!\left(\!\frac{B^4(1\!+\!\delta)^2}{K_m\hat{\omega}_m^{\text{\begin{tiny}min\end{tiny}}}\omega_m^{\text{\begin{tiny}min\end{tiny}}}}\!\!\right)^{\!2}\!\!\!P_{m}}$,
and $\zeta\triangleq\!\sqrt{\sum_{m\in\mathcal{M}}P_{m}}$.
\end{lem}
\textit{Proof:} The proofs of (\ref{BDD}) and (\ref{BDDhat}) follow from
(\ref{lm3-1}). To prove (\ref{BDDDhat}), we omit time index $t$ for notation
simplicity. By the definition of $\mathbf{D}$ and $\hat{\mathbf{D}}$, we
have
\begin{align}
        \Vert \mathbf{D}-\hat{\mathbf{D}} \Vert_F^2=\sum_{m\in\mathcal{M}}\Vert\mathbf{H}_{m}\mathbf{W}_{m}-\hat{\mathbf{H}}_{m}\hat{\mathbf{W}}_{m}\Vert_F^2.\label{lmD-1}
\end{align}
Using the MRT precoding in (\ref{MRT}) for $m\in\mathcal{M}_\text{\begin{tiny}MRT\end{tiny}}$
we have
\begin{align}
         &\Vert\mathbf{H}_{m}\mathbf{W}_{m}^{\text{\begin{tiny}MRT\end{tiny}}}-\hat{\mathbf{H}}_{m}\hat{\mathbf{W}}_{m}^{\text{\begin{tiny}MRT\end{tiny}}}\Vert_F\notag\\
         &=\sqrt{P_{m}}\left\Vert\frac{\mathbf{H}_{m}\mathbf{H}_{m}^H}{\Vert\mathbf{H}_{m}\Vert_F}-\frac{\hat{\mathbf{H}}_{m}\hat{\mathbf{H}}_{m}^H}{\Vert\hat{\mathbf{H}}_{m}\Vert_F}\right\Vert_F\notag\\
         &=\sqrt{P_m}\left\Vert\frac{\mathbf{H}_{m}\mathbf{H}_{m}^H}{\Vert\mathbf{H}_{m}\Vert_F}-\frac{(\mathbf{H}_{m}-\tilde{\mathbf{H}}_{m})(\mathbf{H}_{m}^H-\tilde{\mathbf{H}}_{m}^H)}{\Vert\mathbf{H}_{m}-\tilde{\mathbf{H}}_{m}\Vert_F}\right\Vert_F\notag\\
         &=\sqrt{P_m}\left\Vert\left(\frac{\mathbf{H}_{m}\mathbf{H}_{m}^H}{\Vert\mathbf{H}_{m}\Vert_F}-\frac{\mathbf{H}_{m}\mathbf{H}_{m}^H}{\Vert\mathbf{H}_{m}-\tilde{\mathbf{H}}_{m}\Vert_F}\right)\right.\notag\\
         &\quad\left.-\frac{\tilde{\mathbf{H}}_{m}\tilde{\mathbf{H}}_{m}^H-2\Re\{\tilde{\mathbf{H}}_{m}\mathbf{H}_{m}^H\}}{\Vert\mathbf{H}_{m}-\tilde{\mathbf{H}}_{m}\Vert_F}\right\Vert_F\notag\\
         &\leq\sqrt{P_{m}}\left[ \frac{\Vert\mathbf{H}_{m}\mathbf{H}_{m}^H\Vert_F}{\Vert\mathbf{H}_{m}\Vert_F}\left(1-\frac{\Vert\mathbf{H}_{m}\Vert_F}{\Vert\mathbf{H}_{m}-\tilde{\mathbf{H}}_{m}\Vert_F}\right)\right.\notag\\
         &\quad\left.+\frac{\Vert\tilde{\mathbf{H}}_{m}\tilde{\mathbf{H}}_{m}^H-2\Re\{\tilde{\mathbf{H}}_{m}\mathbf{H}_{m}^H\}\Vert_F}{\Vert\mathbf{H}_{m}-\tilde{\mathbf{H}}_{m}\Vert_F}\right]\notag\\
         &\stackrel{(a)}{\leq} \sqrt{P_{m}}\left( B\delta+\frac{\Vert\tilde{\mathbf{H}}_{m}\tilde{\mathbf{H}}_{m}^H-2\Re\{\tilde{\mathbf{H}}_{m}\mathbf{H}_{m}^H\}\Vert_F}{\Vert\mathbf{H}_{m}-\tilde{\mathbf{H}}_{m}\Vert_F}\right)\notag\\
         &\stackrel{(b)}{\leq}\sqrt{P_{m}}B\delta\!\left(\!\!1\!+ \!\frac{(2\!+\!\delta)B}{\Vert\hat{\mathbf{H}}_{m}\Vert_F}\!\!\right)\!\leq\!\sqrt{P_{m}}B\delta\!\left(\!\!1\!+\!\frac{(2\!+\!\delta)B}{\hat{B}_{m}^{\text{\begin{tiny}min\end{tiny}}}}\!\!\right)\!\label{DMRT}
\end{align}
where $(a)$ is because
\begin{align*}
        &\frac{\Vert\mathbf{H}_{m}\mathbf{H}_{m}^H\Vert_F}{\Vert\mathbf{H}_{m}\Vert_F}\left(\!1\!-\!\frac{\Vert\mathbf{H}_{m}\Vert_F}{\Vert\mathbf{H}_{m}\!-\!\tilde{\mathbf{H}}_{m}\Vert_F}\!\right)\\
        &\le\frac{\Vert\mathbf{H}_m\Vert_F^2}{\Vert\mathbf{H}_m\Vert_F}\left(1-\frac{\Vert\mathbf{H}_{m}\Vert_F}{\Vert\tilde{\mathbf{H}}_{m}\Vert_F\!+\!\Vert\mathbf{H}_{m}\Vert_F}\right)\!\leq{\!B}\!\left(\!1\!-\!\frac{1}{1\!+\!\delta}\!\right)\!\leq\!B\delta,
\end{align*}
in which we use $\Vert\mathbf{H}_m\Vert_F\!\le\!{B}$
and $\Vert\tilde{\mathbf{H}}_m\Vert_F\!\le\!{B\delta}$ from (\ref{eq:asm1})
and  (\ref{eq:asm2}), respectively;
and $(b)$ is because
\begin{align*}
        &\Vert\tilde{\mathbf{H}}_{m}\tilde{\mathbf{H}}_{m}^H-2\Re\{\tilde{\mathbf{H}}_{m}\mathbf{H}_{m}^H\}\Vert_F\leq\Vert\tilde{\mathbf{H}}_{m}\tilde{\mathbf{H}}_{m}^H\Vert_F+2\Vert\tilde{\mathbf{H}}_{m}\mathbf{H}_{m}^H\Vert_F\\
        &\leq \!\Vert\tilde{\mathbf{H}}_{m}\Vert_F^2\!+\!2\Vert\tilde{\mathbf{H}}_{m}\Vert_F\Vert\mathbf{H}_{m}\Vert_F\leq\!B^2\delta^2\!+\!2B^2\delta\leq\!(2\!+\!\delta)B^2\delta.
\end{align*}

With ZF precoding in (\ref{ZF}) for $m\in\mathcal{M}\backslash\mathcal{M}_\text{\begin{tiny}MRT\end{tiny}}$,
we have
\begin{align}
        &\Vert\mathbf{H}_{m}\mathbf{W}_{m}^{\text{\begin{tiny}ZF\end{tiny}}}-\hat{\mathbf{H}}_{m}\hat{\mathbf{W}}_{m}^{\text{\begin{tiny}ZF\end{tiny}}}\Vert_F\notag\\
        &=\sqrt{P_m}\left\Vert\frac{\mathbf{I}}{\sqrt{\tr\{(\mathbf{H}_m\mathbf{H}_m^H)^{-1}\}}}-\frac{\mathbf{I}}{\sqrt{\tr\{(\hat{\mathbf{H}}_m\hat{\mathbf{H}}_m^H)^{-1}\}}}\right\Vert_F\notag\\
        &\stackrel{(a)}{\leq}\!\sqrt{P_mK_m}\frac{\left|\sqrt{\sum_{i\in\mathcal{K}_m}\!\hat{\omega}_{m,i}^{-1}}-\!\sqrt{\sum_{i\in\mathcal{K}_m}\!\omega_{m,i}^{-1}}\right|}{\frac{K_{m}}{B^{2}(1+\delta)}}\notag\\
        &=\!\sqrt{\frac{P_m}{K_m}}B^{2}(1\!+\!\delta)\frac{\left|\sum_{i\in\mathcal{K}_m}\left(\hat{\omega}_{m,i}^{-1}-\omega_{m,i}^{-1}\right)\right|}{\sqrt{\sum_{i\in\mathcal{K}_m}\!\hat{\omega}_{m,i}^{-1}}+\sqrt{\sum_{i\in\mathcal{K}_m}\!\omega_{m,i}^{-1}}}\notag\\
        &\stackrel{(b)}{\leq}\!\sqrt{\frac{P_m}{K_m}}B^{2}(1+\delta)\frac{B^{2}(2+\delta)\delta}{\hat{\omega}_m^{\text{\begin{tiny}min\end{tiny}}}\omega_m^{\text{\begin{tiny}min\end{tiny}}}\frac{(2+\delta)\sqrt{K_m}}{B(1+\delta)}}\leq\frac{\sqrt{P_m}B^{5}(1+\delta)^2\delta}{K_{m}\hat{\omega}_m^{\text{\begin{tiny}min\end{tiny}}}\omega_m^{\text{\begin{tiny}min\end{tiny}}}}\label{DZF}
\end{align}
where $(a)$ follows from $\hat{\omega}_{m,i}\le\Vert\hat{\mathbf{H}}_m\Vert_F^2,\forall{i}\in\mathcal{K}_m$,
and $\Vert\hat{\mathbf{H}}\Vert_F\le{B(1+\delta)}$ in (\ref{Hhat}), such
that $\tr\{(\hat{\mathbf{H}}_m\hat{\mathbf{H}}_m^H)^{-1}\}=\sum_{i\in\mathcal{K}_m}\hat{\omega}_{m,i}^{-1}\geq\frac{K_{m}}{B^{2}(1+\delta)^2}$,
and similarly $\tr\{(\mathbf{H}_m\mathbf{H}_m^H)^{-1}\}=\sum_{i\in\mathcal{K}_m}\omega_{m,i}^{-1}\ge\frac{K_m}{B^2}$
for accurate CSI; $(b)$ is because
\begin{align*}
        &\sqrt{\sum_{i\in\mathcal{K}_m}\hat{\omega}_{m,i}^{-1}}+\!\sqrt{\sum_{i\in\mathcal{K}_m}\omega_{m,i}^{-1}}\ge\sqrt{\frac{K_{m}}{\Vert\hat{\mathbf{H}}_m\Vert_F^2}}+\sqrt{\frac{K_{m}}{\Vert\mathbf{H}_m\Vert_F^2}}\notag\\
        &\geq\sqrt{\frac{K_{m}}{B^{2}(1+\delta)^2}}+\sqrt{\frac{K_{m}}{B^{2}}}=\frac{(2+\delta)\sqrt{K_m}}{B(1+\delta)},\notag\\
        &\left|\sum_{i\in\mathcal{K}_m}\left(\hat{\omega}_{m,i}^{-1}-\omega_{m,i}^{-1}\right)\right|\notag\\
        &\leq\frac{\left|\sum_{i\in\mathcal{K}_m}\left(\omega_{m,i}-\hat{\omega}_{m,i}\right)\right|}{\hat{\omega}_m^{\text{\begin{tiny}min\end{tiny}}}\omega_m^{\text{\begin{tiny}min\end{tiny}}}}=\frac{\left|\Vert\mathbf{H}_m\Vert_F^2-\Vert\hat{\mathbf{H}}_m\Vert_F^2\right|}{\hat{\omega}_m^{\text{\begin{tiny}min\end{tiny}}}\omega_m^{\text{\begin{tiny}min\end{tiny}}}}\notag\\
        &=\frac{(\Vert\mathbf{H}_m\Vert_F+\Vert\hat{\mathbf{H}}_m\Vert_F)\left|\Vert\mathbf{H}_m\Vert_F-\Vert\hat{\mathbf{H}}_m\Vert_F\right|}{\hat{\omega}_m^{\text{\begin{tiny}min\end{tiny}}}\omega_m^{\text{\begin{tiny}min\end{tiny}}}}\leq\frac{B^{2}(2+\delta)\delta}{\hat{\omega}_m^{\text{\begin{tiny}min\end{tiny}}}\omega_m^{\text{\begin{tiny}min\end{tiny}}}}
\end{align*}
where we apply $\Vert\mathbf{\mathbf{H}}_m\Vert_F\le{B}$, $\Vert\hat{\mathbf{H}}_m\Vert_F\le{B}(1+\delta)$,
and $|\Vert\mathbf{H}_m\Vert_F-\Vert\hat{\mathbf{H}}_m\Vert_F|\le\Vert\mathbf{H}_m-\hat{\mathbf{H}}_m\Vert_F\le{B}\delta$
resulting from (\ref{eq:asm1}) and (\ref{eq:asm2}) to the last step.

Applying inequalities (\ref{DMRT}) and (\ref{DZF}) to the respective terms
in the RHS of (\ref{lmD-1}) yields (\ref{BDDDhat}).
\endIEEEproof

For channel state $\mathbf{H}(t)$ being i.i.d. over time, there exists a
stationary randomized optimal precoding solution $\mathbf{V}^{\text{\begin{tiny}opt\end{tiny}}}(t)$
to \textbf{P1}, which depends only on the (unknown) distribution of  $\mathbf{H}(t)$
and achieves the minimum objective value of \textbf{P1}, defined in Theorem
\ref{lmHBD} \cite{Neely}. Define $\phi(\mathbf{H}(t),\mathbf{V}(t),\mathbf{D}(t))\triangleq{U}\Vert\mathbf{H}(t)\mathbf{V}(t)-\mathbf{D}(t)\Vert_F^2+Z(t)\Vert\mathbf{V}(t)\Vert_F^2$.
Note that $\phi(\hat{\mathbf{H}}(t),\hat{\mathbf{V}}(t),\hat{\mathbf{D}}(t))$
is the objective function in \textbf{P2}. Using Lemma \ref{Dbd}, for a given
CSI inaccuracy $\delta$ in (\ref{eq:asm2}), we now bound $\phi(\mathbf{H}(t),\hat{\mathbf{V}}^{\star}(t),\mathbf{D}(t))-\phi(\mathbf{H}(t),\mathbf{V}^{\text{\begin{tiny}opt\end{tiny}}}(t),\mathbf{\mathbf{D}}(t))$,
where the first term is the objective value of \textbf{P2} under the optimal
solution $\hat{\mathbf{V}}^{\star}(t)$ to \textbf{P2} obtained based on the
inaccurate channel state $\hat{\mathbf{H}}(t)$, and the second term is the
objective value of \textbf{P2} by using the optimal solution $\mathbf{V}^{\text{\begin{tiny}opt\end{tiny}}}(t)$
to \textbf{P1} obtained based on the accurate channel state $\mathbf{H}(t)$.

\begin{lem}\label{lmrho}
At each time $t$, the following holds
\begin{align}
        \phi(\mathbf{H}(t),\hat{\mathbf{V}}^{\star}(t),\mathbf{D}(t))-\phi(\mathbf{H}(t),\mathbf{V}^{\text{\begin{tiny}opt\end{tiny}}}(t),\mathbf{D}(t))\leq{U}\varphi\label{P2BD}
\end{align}
where\begin{align*}
        \varphi\triangleq2\!\left[(2\!+\!\delta)(P_{\text{max}}\!+\!\zeta\eta)\!+\!2(\zeta(1\!+\!\delta)\!+\!\eta)\sqrt{\!P_{\text{max}}}\right]\!B^{2}\delta\!=\!O(\delta).
\end{align*}
\end{lem}
\textit{Proof:} We omit time index $t$ in the proof. The proof of (\ref{P2BD})
consists of five steps as follows.

\textit{Step 1:} 
Note that $\phi(\mathbf{H},\hat{\mathbf{V}}^{\star},\mathbf{D})$ is convex
with respect to (w.r.t.) $\mathbf{D}$. By the first-order condition for a
convex function~\cite{GD}, we have
\begin{align*}
        &\phi(\mathbf{H},\hat{\mathbf{V}}^{\star},\mathbf{D})-\phi(\mathbf{H},\hat{\mathbf{V}}^{\star},\hat{\mathbf{D}})\notag\\
        &\leq-2\Re\{\tr\{\nabla_{\mathbf{\mathbf{D}}^{*}}\phi(\mathbf{H},\hat{\mathbf{V}}^{\star},\mathbf{D})^H(\hat{\mathbf{D}}-\mathbf{D})\}\}\notag\\
        &\stackrel{(a)}{=} 2U\Re\{\tr\{ (\mathbf{H}\hat{\mathbf{V}}^{\star}-\mathbf{D})^H(\hat{\mathbf{D}}-\mathbf{D})\}\}\notag\\
        &\leq2U|\tr\{(\mathbf{H}\hat{\mathbf{V}}^{\star}-\mathbf{D})^H(\hat{\mathbf{D}}-\mathbf{D})\}|\notag\\
        &\leq 2U(\Vert \mathbf{H}\Vert_F\Vert\hat{\mathbf{V}}^{\star}\Vert_F+\Vert\mathbf{D}\Vert_F)\Vert\hat{\mathbf{D}}-\mathbf{D}\Vert_F\notag\\
        &\stackrel{(b)}{\leq}{2}U(\sqrt{P_{\text{max}}}+\zeta)\eta{B}^{2}\delta
\end{align*}
where $(a)$ follows from $\nabla_{\mathbf{D}^{*}} \phi(\mathbf{H},\hat{\mathbf{V}}^{\star},\mathbf{\mathbf{D}})=-U(\mathbf{H}\hat{\mathbf{V}}^{\star}-\mathbf{D})$,
and $(b)$ follow from (\ref{eq:asm1}), (\ref{sl2}),
(\ref{BDD}), and (\ref{BDDDhat}).

\textit{Step 2:} 
By the first-order condition for the convex function $\phi(\mathbf{H},\hat{\mathbf{V}}^{\star},\hat{\mathbf{D}})$
w.r.t. $\mathbf{H}$, we have
\begin{align*}
        &\phi(\mathbf{H},\hat{\mathbf{V}}^{\star},\hat{\mathbf{D}})-\phi(\hat{\mathbf{H}},\hat{\mathbf{V}}^{\star},\hat{\mathbf{D}})\notag\\
        &\leq-{2}\Re\{\tr\{\nabla_{\mathbf{H}^*}\phi(\mathbf{H},\hat{\mathbf{V}}^{\star},\hat{\mathbf{D}})^H(\hat{\mathbf{H}}-\mathbf{H})\}\}\notag\\
        &\stackrel{(a)}{=}2U\Re\{\tr\{(\mathbf{H}\hat{\mathbf{V}}^{\star}\hat{\mathbf{V}}^{\star{H}}-\hat{\mathbf{D}}\hat{\mathbf{V}}^{\star{H}})^H(\mathbf{H}-\hat{\mathbf{H}})\}\}\notag\\
        &\leq{2U}|\tr\{(\mathbf{H}\hat{\mathbf{V}}^{\star}\hat{\mathbf{V}}^{\star{H}}-\hat{\mathbf{D}}\hat{\mathbf{V}}^{\star{H}})^H(\mathbf{H}-\hat{\mathbf{H}})\}|\notag\\
        &\leq 2U( \Vert \mathbf{H}\Vert_F \Vert\mathbf{\hat{\mathbf{V}}}^{\star}\Vert_F+\Vert\hat{\mathbf{D}}\Vert_F)\Vert\hat{\mathbf{V}}^{\star}\Vert_F\Vert\mathbf{H}-\hat{\mathbf{H}}\Vert_F\notag\\
        &\stackrel{(b)}{\leq}2U\left[P_{\text{max}}+\zeta(1+\delta)\sqrt{P_{\text{max}}}\right]B^{2}\delta
\end{align*}
where $(a)$ follows from $\nabla_{\mathbf{H}^{*}} \phi(\mathbf{H},\hat{\mathbf{V}}^{\star},\hat{\mathbf{D}})=U(\mathbf{H}\hat{\mathbf{V}}^{\star}\hat{\mathbf{V}}^{\star{H}}-\hat{\mathbf{D}}\hat{\mathbf{V}}^{\star{H}})$,
and $(b)$ follows from (\ref{eq:asm1}), (\ref{eq:asm2}), (\ref{sl2}), and
(\ref{BDDhat}).

\textit{Step 3:} In Algorithm \ref{alg:1}, $\hat{\mathbf{V}}^{\star}$ is
the optimal precoder that minimizes the objective  $\phi(\hat{\mathbf{H}},\hat{\mathbf{V}},\hat{\mathbf{D}})$
of \textbf{P2} over all precoding policies including $\mathbf{V}^{\text{\begin{tiny}opt\end{tiny}}}$.
It follows that
\begin{align*}
        \phi(\hat{\mathbf{H}},\hat{\mathbf{V}}^{\star},\hat{\mathbf{D}})-\phi(\hat{\mathbf{H}},\mathbf{V}^{\text{\begin{tiny}opt\end{tiny}}},\hat{\mathbf{D}})\leq0.
\end{align*}

\textit{Step 4:} Similarly to Step 2, we have
\begin{align*}
        &\phi(\hat{\mathbf{H}},\mathbf{V}^{\text{\begin{tiny}opt\end{tiny}}},\hat{\mathbf{D}})-\phi(\mathbf{H},\mathbf{V}^{\text{\begin{tiny}opt\end{tiny}}},\hat{\mathbf{D}})\notag\\
        &\le2U(P_{\text{max}}+\zeta\sqrt{P_{\text{max}}})(1+\delta)B^{2}\delta.
\end{align*}

\textit{Step 5:} Similarly to Step 1, we have
\begin{align*}
        &\phi(\mathbf{H},\mathbf{V}^{\text{\begin{tiny}opt\end{tiny}}},\hat{\mathbf{D}})-\phi(\mathbf{H},\mathbf{V}^{\text{\begin{tiny}opt\end{tiny}}},\mathbf{D})\notag\\
        &\leq{2U}\left[\sqrt{P_{\text{max}}}+\zeta(1+\delta)\right]\eta{B}^{2}\delta.
\end{align*}

Summing over Steps 1-5, yields (\ref{P2BD}).\endIEEEproof

\begin{remark}
We point out that, different from \cite{HY}, our proof in Lemma \ref{lmrho}
explicitly considers the two-fold impact of CSI inaccuracy on both
InP and SPs under MIMO WNV.
\end{remark}

Based on Lemma \ref{lmrho}, we next show that with the optimal $\hat{\mathbf{V}}(t)$
to \textbf{P2}, the expected DPP metric averaged over the virtual queue $Z(t)$
under  the accurate channel state $\mathbf{H}(t)$ is upper bounded. 
\begin{lem} \label{lmdpp}
        At each time $t$, we have
\begin{align}
        &\mathbb{E}\{\Delta(t)\}+U\mathbb{E}\{\Vert\mathbf{H}(t)\hat{\mathbf{V}}^{\star}(t)-\mathbf{D}(t)\Vert_F^{2}\}\notag\\
        &\leq{U}\mathbb{E}\{\Vert\mathbf{H}(t)\mathbf{V}^{\text{\begin{tiny}opt\end{tiny}}}(t)-\mathbf{D}(t)\Vert_F^2\}+U\varphi+S\label{eqdpp}
\end{align}
where $\varphi$ is given in Lemma 4 and $S$ is defined below (\ref{DPPUB}).
\end{lem}
\textit{Proof:} From (\ref{eq:lm1-1}) in the proof of Lemma \ref{lmDPPUB},
at each time $t$, the Lyapunov drift $\Delta(t)$ is upper bounded as  $\Delta(t)\leq{Z}(t)(\Vert\hat{\mathbf{V}}^{\star}(t)\Vert_F^2-\bar{P})+S$.
Adding $U\Vert \mathbf{H}(t)\hat{\mathbf{V}}^{\star}(t)-\mathbf{D}(t)\Vert_F^2$
at both sides of the above inequality yields
\begin{align}
        &\Delta(t)+U\Vert \mathbf{H}(t)\hat{\mathbf{V}}^{\star}(t)-\mathbf{D}(t)\Vert_F^2\notag\\
        &\leq{U}\Vert\mathbf{H}(t)\hat{\mathbf{V}}^{\star}(t)-\mathbf{D}(t)\Vert_F^2+Z(t)(\Vert\hat{\mathbf{V}}^{\star}(t)\Vert_F^2-\bar{P})+S\notag\\
        &\stackrel{(a)}{\leq} U\Vert \mathbf{H}(t)\mathbf{V}^{\text{\begin{tiny}opt\end{tiny}}}(t)-\mathbf{D}(t)\Vert_F^2+Z(t)(\Vert\hat{\mathbf{V}}^{\text{\begin{tiny}opt\end{tiny}}}(t)\Vert_F^2-\bar{P})\notag\\
        &\quad+U\varphi+S\label{eq:lm5-2}
\end{align}
where $(a)$ follows from (\ref{P2BD}) in Lemma \ref{lmrho}. Taking
expectations at both sides of (\ref{eq:lm5-2}), we have
\begin{align*}
        &\mathbb{E}\left\{ \Delta(t) \right\} + U \mathbb{E} \{\Vert \mathbf{H}(t)\hat{\mathbf{V}}^{\star}(t)-\mathbf{D}(t)\Vert_F^2\}\\
        &\leq U \mathbb{E}\left\{ \Vert \mathbf{H}(t)\mathbf{V}^{\text{\begin{tiny}opt\end{tiny}}}(t)-\mathbf{D}(t)\Vert_F^2\right\}\!+\!\mathbb{E}\left\{Z(t)(\Vert\mathbf{V}^{\text{\begin{tiny}opt\end{tiny}}}(t)\Vert_F^2-\bar{P})\right\}\notag\\
        &\quad+U\varphi+S\\
        &= U \mathbb{E}\{\Vert \mathbf{H}(t)\mathbf{V}^{\text{\begin{tiny}opt\end{tiny}}}(t)-\mathbf{D}(t)\Vert_F^2\}\notag\\
        &\quad+\mathbb{E}\left\{\mathbb{E}\left\{Z(t)(\Vert\mathbf{V}^{\text{\begin{tiny}opt\end{tiny}}}(t)\Vert_F^2-\bar{P})|Z(t)\right\}\right\}+U\varphi+S\\
        &\stackrel{(a)}{\leq} U \mathbb{E}\left\{ \Vert \mathbf{H}(t)\mathbf{V}^{\text{\begin{tiny}opt\end{tiny}}}(t)-\mathbf{D}(t)\Vert_F^2\right\}+U\varphi+S
\end{align*}
where $(a)$ is because  the optimal $\mathbf{V}^{\text{\begin{tiny}opt\end{tiny}}}(t)$
to \textbf{P1} depends only on $\mathbf{H}(t)$ and  is independent of
$Z(t)$. Since $Z(t)\ge0$, it follows that $\mathbb{E}\left\{ Z(t)(\Vert\mathbf{V}^{\text{\begin{tiny}opt\end{tiny}}}(t)\Vert_F^2\!-\!\bar{P})|Z(t)\right\}\!=\!Z(t)\mathbb{E}\{\Vert\mathbf{V}^{\text{\begin{tiny}opt\end{tiny}}}(t)\Vert_F^2-\bar{P}\}\!\leq\!0$.
\endIEEEproof

\begin{remark}
Note that the standard Lyapunov optimization relies on an upper bound analysis
of the DPP metric under the accurate system state \cite{Neely}, which is
the accurate channel state $\mathbf{H}(t)$ in our MIMO WNV problem. Our results
in Lemma~\ref{lmdpp} extends that analysis to inaccurate channel state $\hat{\mathbf{H}}(t)$.
Different from the accurate CSI, the inaccurate CSI causes a two-fold impact
on both the InP and SPs in our MIMO WNV problem, which complicates the analysis.
\end{remark}

Based on Lemmas \ref{lmZ} and \ref{lmdpp} and by Lyapunov optimization techniques
\cite{Neely}, we provide the performance bounds for Algorithm \ref{alg:1}
with imperfect CSI in the following theorem.

\begin{thm} \label{lmHBD}
For any $\epsilon>0$, set $U=\frac{S}{\epsilon}$ in Algorithm~\ref{alg:1}.
Consider $\hat{\mathbf{V}}^{\star}(t)$ produced by Algorithm \ref{alg:1}
based on $\hat{\mathbf{H}}(t)$. For any $T>0$, the following hold regardless
of the distribution of $\mathbf{H}(t)$:
\begin{align}
        &\!\frac{1}{T}\sum_{t=0}^{T-1}\mathbb{E}\left\{\Vert\mathbf{H}(t)\hat{\mathbf{V}}^{\star}(t)-\mathbf{D}(t)\Vert_F^2\right\}\leq\rho^{\text{\begin{tiny}opt\end{tiny}}}+\varphi+\epsilon,\!\label{thm1}\\
        &\!\frac{1}{T} \sum_{t=0}^{T-1}\Vert \hat{\mathbf{V}}^{\star}(t)\Vert_F^2\leq\bar{P}+\frac{SB^2(1+\delta)^2\xi+\epsilon(P_{\text{max}}-\bar{P})}{\epsilon{T}}\!\label{thm2}
\end{align}
where $\rho^{\text{\begin{tiny}opt\end{tiny}}}$ is the minimum objective
value of \textbf{P1} with $\mathbf{H}(t)$, $\varphi=\mathcal{O}(\delta)$
is defined below (\ref{P2BD}), and $\xi$ is defined below (\ref{BDVQ}). 
\end{thm}
\textit{Proof:} We first prove (\ref{thm1}). The long-term time-averaged
expected precoding deviation in the LHS of (\ref{thm1}) is upper bounded
as
\begin{align}
        &\frac{1}{T}\sum_{t=0}^{T-1}\mathbb{E}\{\Vert \mathbf{H}(t)\hat{\mathbf{V}}^{\star}(t)-\mathbf{D}(t)\Vert_F^{2}\}\notag\\
        &\stackrel{(a)}{\leq}\frac{1}{T}\sum_{t=0}^{T-1}\mathbb{E}\{\Vert\mathbf{H}(t)\mathbf{V}^{\text{\begin{tiny}opt\end{tiny}}}(t)-\mathbf{D}(t)\Vert_F^2\}-\frac{1}{UT}\sum_{t=0}^{T-1}\mathbb{E}\{\Delta(t)\}\notag\\
        &\quad+\varphi+\frac{S}{U}\notag\\
        &\stackrel{(b)}{\leq} \rho^{\text{\begin{tiny}opt\end{tiny}}}-\frac{1}{2TU}(\mathbb{E}\{{Z}^2(T)\}-\mathbb{E}\{{Z}^2(0)\})+\varphi+\frac{S}{U}\notag\\
        &\stackrel{(c)}{\leq}\rho^{\text{\begin{tiny}opt\end{tiny}}}+\varphi+\frac{S}{U}\label{eq:lm6-1}
\end{align}
where $(a)$ is obtained by summing the terms in (\ref{eqdpp}) over time $t$
from $0$ to $T-1$, dividing them by $UT$, and rearranging them; $(b)$ follows
from $\sum_{t=0}^{T-1}\mathbb{E}\{\Delta(t)\}=\sum_{t=0}^{T-1}\frac{1}{2}\mathbb{E}\{{Z}^2(t+1)-{Z}^2(t)\}=\frac{1}{2}\mathbb{E}\{{Z}^2(T)\}-\mathbb{E}\{{Z}^2(0)\}$;
$(c)$ is because $Z(t)\ge0,\forall{t}$, and we set the initial value ${Z}(0)=0$.
Finally, substituting  $U=\frac{S}{\epsilon}$ into (\ref{eq:lm6-1}) yields
(\ref{thm1}).

We now prove (\ref{thm2}). From the virtual queue dynamics in
(\ref{eq:VQ}), we have $Z(t+1)\ge Z(t)+\Vert\hat{\mathbf{V}}^\star(t)\Vert_F^2-\bar{P},~\forall{t}$.
Rearranging terms of the above inequality, we have $\Vert\hat{\mathbf{V}}^\star(t)\Vert_F^2\le\bar{P}+Z(t+1)-Z(t),~\forall{t}$.
Summing both sides of the above inequality over $t$ from $0$ to $T-1$ and
then dividing by $T$ yields 
\begin{align*}
\frac{1}{T}\sum_{t=0}^{T-1}\Vert\hat{\mathbf{V}}^\star(t)\Vert_F^2\leq\bar{P}+\frac{Z(T)-Z(0)}{T}=\bar{P}+\frac{Z(T)}{T}.
\end{align*}
Substituting the upper bound of $Z(t)$ in (\ref{BDVQ}), and $U=\frac{S}{\epsilon}$
into the above inequality, we have (\ref{thm2}).\endIEEEproof
      
Theorem \ref{lmHBD} provides an upper bound on the objective value of \textbf{P1}
in (\ref{thm1}) achieved by Algorithm~\ref{alg:1}, \ie the time-averaged
expected precoding deviation using $\hat{\mathbf{V}}(t)$ from the virtualization
demand $\hat{\mathbf{D}}(t)$ under inaccurate CSI. It indicates that, for
any given $T$, the performance of Algorithm~\ref{alg:1} using  inaccurate
channel state $\hat{\mathbf{H}}(t)$ can be arbitrarily close to the minimum
deviation $\rho^{\text{\begin{tiny}opt\end{tiny}}}$ achieved using accurate
channel state $\mathbf{H}(t)$ plus a constant term as a function of CSI inaccuracy
$O(\delta)$. The performance gap $\epsilon$ is a controllable parameter by
our design and can be set arbitrarily small. Note that this analysis is different
from the standard $(\epsilon,\frac{1}{\epsilon})$ trade-off in Lyapunov optimization
with accurate system state information \cite{Neely}. Furthermore, (\ref{thm2})
provides a bound on the average transmit power over $T$ time slots. It indicates
that for any $T\geq\frac{1}{\epsilon^2}$, Algorithm \ref{alg:1} guarantees
that the deviation of the average power from the long-term average transmit
power limit $\bar{P}$ is within $O(\epsilon)$. In particular, as $T\to\infty$,
(\ref{thm2}) becomes the long-term average transmit power constraint (\ref{VM2}),
and it implies Algorithm~\ref{alg:1} satisfies (\ref{VM2}) asymptotically.

\section{Online Coordinated Multi-Cell MIMO WNV}
\label{sec:multi-cell}

In this section, we extend the online MIMO WNV solution of the single-cell
case to the multi-cell scenario. With multiple cells, the level of coordination
and how to perform distributed implementation are two critical issues. Existing
works focus on offline coordinated precoding designs for non-virtualized
networks \cite{W.Yu-CoOD}-\nocite{X.Wang-CoOD}\cite{T.Queck-CoOD}. In contrast,
we propose an online multi-cell coordinated precoding scheme for network
virtualization. The proposed scheme naturally leads to a \textit{fully distributed}
implementation at each cell.

\subsection{Multi-Cell Spatial Virtualization}

Consider a virtualized multi-cell MIMO network where an InP performs virtualization
at each cell for multiple SPs. The subscribing-user sets of different SPs
are disjoint, and each user is only served by its serving cell. For interference
mitigation, multiple cells are coordinated via transmit precoding, with no
CSI exchange across cells.

Specifically, consider an InP that performs virtualization among $C$ cells.
Let $\mathcal{C}=\{1,\dots,C\}$. The BS $c\in\mathcal{C}$ has $N^c$ antennas.
The total number of antennas in the network is $N=\sum_{c\in\mathcal{C}}N^c$.
Each SP $m\in\mathcal{M}$ has $K_m^c$ users in cell $c$. The total number
of users in cell $c$ is $K^c=\sum_{m\in\mathcal{M}}K_m^c$, and that in the
network is $K=\sum_{c\in\mathcal{C}}K^c$. Let $\mathcal{K}_m^c=\{1,\dots,K_m^c\}$
and $\mathcal{K}^c=\{1,\dots,K^c\}$.

Let $\mathbf{H}_m^{cl}(t)\in\mathbb{C}^{K_m^c\times{N}^l}$ denote the  channel
state between SP~$m$'s subscribing users in cell $c$ and BS~$l$. For ease
of exposition, we first illustrate idealized multi-cell spatial virtualization
where the CSI estimation if perfect. At each time $t$, at each BS $c$, the
InP shares the local channel state $\mathbf{H}_m^{cc}(t)$ with SP $m$ and
allocates a transmission power $P_m^c$ to the SP. Using $\mathbf{H}_m^{cc}(t)$,
SP $m$ designs its precoding matrix $\mathbf{W}_m^c(t)\in\mathbb{C}^{N^c\times{K}_m^c}$,
subject to the transmission power limit $\Vert\mathbf{W}_m^c(t)\Vert_F^2\leq{P}_m^c$.
SP $m$ then sends $\mathbf{W}_m^c(t)$ to the InP as its virtual precoding
matrix. For SP $m$, with $\mathbf{W}_m^c(t)$, the \textit{desired} received
signal vector (noiseless) $\tilde{\mathbf{y}}_m^c$ (at $K_m^c$ users) is
given~by 
\begin{align*}
        \tilde{\mathbf{y}}_m^c(t)=\mathbf{H}_m^{cc}(t)\mathbf{W}_m^c(t)\mathbf{x}_m^c(t)
\end{align*}
where $\mathbf{x}_m^c(t)$ is the symbol vector to SP $m$'s users. Define
 $\tilde{\mathbf{y}}^c(t)\triangleq[\tilde{\mathbf{y}}_1^{cH}(t),\dots,\tilde{\mathbf{y}}_M^{cH}(t)]^H$
as the desired received signal vector at all $K^c$ users in cell $c$, we
have
\begin{align*}
        \tilde{\mathbf{y}}^c(t)= \mathbf{D}^c(t)\mathbf{x}^c(t)
\end{align*} 
where $\mathbf{D}^c(t)\triangleq\blkdiag\{\mathbf{H}_1^{cc}(t)\mathbf{W}_1^c(t),\dots,\mathbf{H}_M^{cc}(t)\mathbf{W}_M^c(t)\}\in\mathbb{C}^{K^c\times{K}^c}$
is the virtualization demand from all SPs in cell $c$ and $\mathbf{x}^c(t)\triangleq[\mathbf{x}_1^{cH}(t),\dots,\mathbf{x}_M^{cH}(t)]^H$.
Let the desired received signal vector at all $K$ users in the network be
$\tilde{\mathbf{y}}'(t)\triangleq[\tilde{\mathbf{y}}^{1H}(t),\dots,\tilde{\mathbf{y}}^{CH}(t)]^H$.
We have $\tilde{\mathbf{y}}'(t)=\mathbf{D}'(t)\mathbf{x}'(t)$, where $\mathbf{D}'(t)\triangleq\blkdiag\{\mathbf{D}^1(t),\dots,\mathbf{D}^C(t)\}$
is the global virtualization demand and $\mathbf{x}'(t)\triangleq[\mathbf{x}^{1H}(t),\dots,\mathbf{x}^{CH}(t)]^H$.

The InP virtualizes the BSs to meet the SPs' virtualization service demands.
Let $\mathbf{H}^{cl}(t)\triangleq[\mathbf{H}_{1}^{clH}(t),\dots,\mathbf{H}_{M}^{clH}(t)]^H\in\mathbb{C}^{K^c\times{N}^l}$
denote the channel state between the users in cell $c$ and the BS $l$. In
cell $c$, based only on the local  channel state $\mathbf{H}^c(t)\triangleq[\mathbf{H}^{1cH}(t),\dots,\mathbf{H}^{CcH}(t)]^H\in\mathbb{C}^{K\times{N}^c}$
from all users to BS $c$, the InP designs the actual downlink precoding matrix
$\mathbf{V}^c(t)\triangleq[\mathbf{V}_1^c(t),\dots,\mathbf{V}_M^c(t)]\in\mathbb{C}^{N^c\times{K}^c}$
to serve the $K^c$ users in cell $c$, where $\mathbf{V}_m^c(t)\in\mathbb{C}^{N^c\times{K}_m^c}$
is the precoding matrix designed for SP $m$. The \textit{actual} received
signal vector (noiseless) $\mathbf{y}_m^c(t)$ at the $K_m^c$ users is given
by
\begin{align*}
        \mathbf{y}_m^c(t)&=\mathbf{H}_m^{cc}(t)\mathbf{V}_m^c(t)\mathbf{x}_m^c(t)+\sum_{i\in\mathcal{M},i\neq{m}}\mathbf{H}_m^{cc}\mathbf{V}_i^c(t)\mathbf{x}_i^c(t)\notag\\
        &\quad+\sum_{l\in\mathcal{C},l\neq c}\sum_{j\in\mathcal{M}}\mathbf{H}_m^{cl}(t)\mathbf{V}_j^l(t)\mathbf{x}_j^l(t)
\end{align*}
where the second term is the inter-SP interference from the other SPs' users
in the same cell, and the third term is the inter-cell interference from
users in other cells. The actual received signal vector $\mathbf{y}^c(t)\triangleq[\mathbf{y}_1^{cH}(t),\dots,\mathbf{y}_M^{cH}(t)]^H$
at all $K^c$ users in cell $c$ is given by
\begin{align*}
        \mathbf{y}^c(t)=\mathbf{H}^{cc}(t)\mathbf{V}^c(t)\mathbf{x}^c(t)+\sum_{l\in\mathcal{C},l\neq{c}}\mathbf{H}^{cl}(t)\mathbf{V}^l(t)\mathbf{x}^l(t).
\end{align*}
Let $\mathbf{y}'(t)\triangleq[\mathbf{y}^{1H}(t),\dots,\mathbf{y}^{CH}(t)]^H$
be the actual received signal vector at all $K$ users. We have  $\mathbf{y}'(t)=\mathbf{H}'(t)\mathbf{V}'(t)\mathbf{x}'(t)$,
where $\mathbf{H}'(t)\triangleq[\mathbf{H}^1,\dots,\mathbf{H}^C]$ is the
global channel state and $\mathbf{V}'(t)\triangleq\blkdiag\{\mathbf{V}^1(t),\dots,\mathbf{V}^C(t)\}$
is the InP's actual global precoding matrix.

\subsection{Coordinated Precoding Virtualization Formulation}

Note that each SP $m$ in each cell $c$ designs its virtual precoding matrix
$\mathbf{W}_m^c(t)$ without considering either inter-SP or inter-cell interference.
Therefore, the InP needs to intelligently design the actual global precoding
matrix $\mathbf{V}'(t)$ to mitigate both inter-SP and inter-cell interference,
to meet the virtualization demand $\mathbf{D}'(t)$ from the SPs. The expected
deviation of the actual received signal vector at all $K$ users from the
desired one is given by
\begin{align}
        &\mathbb{E}\{\Vert\mathbf{y}'(t)-\tilde{\mathbf{y}}'(t)\Vert_F^2\}=\mathbb{E}\{\Vert\mathbf{H}'(t)\mathbf{V}'(t)-\mathbf{D}'(t)\Vert_F^2\}\notag\\
        &=\mathbb{E}\left\{\sum_{c\in\mathcal{C}}\Vert\mathbf{H}^c(t)\mathbf{V}^c(t)-\mathbf{G}^c(t)\Vert_F^2\right\}\label{eq:Cobj}
\end{align}
where ${\mathbf{G}^c}(t)\triangleq[\mathbf{0},\dots,\mathbf{D}^{cH}(t),\dots,\mathbf{0}]^H\in\mathbb{C}^{K\times{K}^c}$.

Similar to the single-cell MIMO virtualization problem $\textbf{P1}$, the
online multi-cell coordinated precoding virtualization problem is formulated
as follows:
\begin{align}
        \!\!\!\textbf{P3}:~\min_{\{\mathbf{V}'(t)\}}~&\lim_{T\to\infty}\frac{1}{T}\sum_{t=0}^{T-1}\mathbb{E}\{\Vert\mathbf{H}'(t)\mathbf{V}'(t)-\mathbf{D}'(t)\Vert_F^2\}\notag\\
        \text{s.t.}~~~~ &\lim_{T\to\infty}\frac{1}{T}\sum_{t=0}^{T-1}\mathbb{E}\{\Vert\mathbf{V}^c(t)\Vert_F^2\}\leq\bar{P}^c,~\forall{c}\in\mathcal{C},\!\!\label{CVM2}\\
        &\Vert\mathbf{V}^c(t)\Vert_F^2 \leq P_{\text{max}}^c,~\forall{c}\in\mathcal{C}\label{CVM3}
\end{align}
where $\bar{P}^c$ and $P_{\text{max}}^c$ are the average and the maximum
transmit power limits at the BS in cell $c$, respectively. We assume $\bar{P}^c\leq{P}_{\text{max}}^c,\forall{c}\in\mathcal{C}$.

With the global CSI estimate $\hat{\mathbf{H}}'(t)$ available at each time
$t$, each SP $m$ only has the imperfect local CSI $\hat{\mathbf{H}}_m^{cc}(t)$
provided by the InP to design its virtual precoding matrix, denoted by $\hat{\mathbf{W}}_m^c(t)$.
As a result, the InP receives an inaccurate virtualization demand from $C$
cells $\hat{\mathbf{D}}'(t)\triangleq\blkdiag\{\hat{\mathbf{D}}^1(t),\dots,\hat{\mathbf{D}}^C(t)\}$,
where $\hat{\mathbf{D}}^c(t)\triangleq\blkdiag\{\hat{\mathbf{H}}_1^{cc}(t)\hat{\mathbf{W}}_1^c(t),\dots,\hat{\mathbf{H}}_M^{cc}(t)\hat{\mathbf{W}}_M^c(t)\}$
is the inaccurate virtualization demand from cell $c$. Based on $\hat{\mathbf{H}}'(t)$
and $\hat{\mathbf{D}}'(t)$, the InP designs the actual global precoding
matrix, defined by $\hat{\mathbf{V}}'(t)\triangleq\blkdiag\{\hat{\mathbf{V}}^1(t),\dots,\hat{\mathbf{V}}^C(t)\}$.
In the next subsection, we develop an online multi-cell coordinated MIMO
WNV algorithm based on $\hat{\mathbf{H}}'(t)$ and $\hat{\mathbf{D}}'(t)$
for a coordinated precoding solution $\{\hat{\mathbf{V}}'(t)\}$ to \textbf{P3}.

\subsection{Online Multi-Cell Coordinated MIMO WNV Algorithm}

We extend the online approach developed in the single-cell case to design
an online algorithm to solve \textbf{P3}. We introduce a virtual queue vector
$\mathbf{Z}(t)\triangleq[Z^1(t),\dots,Z^C(t)]^T$. Similar to~(\ref{eq:VQ}),
$Z^c(t)$ is for the per-cell long-term average power constraint (\ref{CVM2})
with the updating rule given~by
\begin{align}
        \!\!Z^c(t+1) = \max\{Z^c(t)+ \Vert \hat{\mathbf{V}}^c(t) \Vert_F^2-\bar{P}^c,0\},~\forall{c}\in\mathcal{C}.\!\!\label{eq:CVQ}
\end{align}
The quadratic Lyapunov function is given by $L(t)=\frac{1}{2} \Vert\mathbf{Z}(t)\Vert_2^2$
and the corresponding Lyapunov drift at time $t$ is given by $\Delta(t)=
L(t+1)-L(t)$. Similar to the single-cell case in Section~\ref{sec:Online
Optimization Formulation}, solving \textbf{P3} can be converted to  minimizing
a DPP metric defined as $\mathbb{E}\{\Delta(t)|\mathbf{Z}(t)\}+U\mathbb{E}\{\hat{\rho}'(t)|\mathbf{Z}(t)\}$,
where $\hat{\rho}'(t)\triangleq\Vert\hat{\mathbf{H}}'(t)\hat{\mathbf{V}}'(t)-\hat{\mathbf{D}}'(t)\Vert_F^2$
and $U>0$ provides the weight between the two terms. We provide an upper
bound for the DPP metric in the following lemma.
\begin{lem}\label{lmDPPUBC}
At each time $t$, for any precoding design of $\hat{\mathbf{V}}'(t)$, the
DPP metric has the following upper bound for all $\mathbf{Z}(t)$ and $U>0$
\begin{align*}
        &\mathbb{E}\{\Delta(t)|\mathbf{Z}(t)\}+U\mathbb{E}\{\hat{\rho}'(t)|\mathbf{Z}(t)\}\notag\\
        &\le{\!S'}\!+\!U\mathbb{E}\left\{\hat{\rho}'(t)|\mathbf{Z}(t)\right\}\!+\!\mathbb{E\!}\left\{\!\sum_{c\in\mathcal{C}}\!Z^c(t)(\Vert\hat{\mathbf{V}}^c(t)\Vert_F^2\!-\!\bar{P}^c)|\mathbf{Z}(t)\!\right\}
\end{align*} where $S'\triangleq\frac{1}{2}\sum_{c\in\mathcal{C}}\max\{(P_{\text{max}}^c-\bar{P}^c)^2,\bar{P}^{c2}\}$.
\end{lem}
\textit{Proof:} See Appendix \ref{APP:lmDPPUBC}.\endIEEEproof

Using the upper bound in Lemma \ref{lmDPPUBC}, with the similar arguments
leading to \textbf{P2}, we have the following per-slot  coordinated precoding
design optimization problem:
\begin{align}
\!\!\!\textbf{P4}:~\min_{\hat{\mathbf{V}}(t)} ~ &U\Vert \hat{\mathbf{H}}'(t)\hat{\mathbf{V}}'(t)-\hat{\mathbf{D}}'(t)\Vert_F^2+\sum_{c\in\mathcal{C}}Z^c(t)\Vert\hat{\mathbf{V}}^c(t)\Vert_F^2\notag\\
\quad\text{s.t.}~~&\Vert\hat{\mathbf{V}}^c(t)\Vert_F^2\leq{P}_{\text{max}}^c,\quad\forall{c}\in\mathcal{C}.\!\!\label{Csl2}
\end{align}
\textbf{P4} can be decomposed into $C$ subproblems, each corresponds to a
local precoding design optimization problem for cell $c$, given by
\begin{align}
        \textbf{P5}:~\min_{\hat{\mathbf{V}}^c(t)} ~ &U\Vert \hat{\mathbf{H}}^c(t)\hat{\mathbf{V}}^c(t)-\hat{\mathbf{G}}^c(t)\Vert_F^2+Z^c\Vert\hat{\mathbf{V}}^c(t)\Vert_F^2\notag\\
        \qquad\text{s.t.}\quad&\Vert\hat{\mathbf{V}}^c(t)\Vert_F^2\leq{P}_{\text{max}}^c\label{Cslc2}
\end{align}
where $\hat{\mathbf{G}}^c(t)\triangleq[\mathbf{0},\dots,\hat{\mathbf{D}}^{cH}(t),\dots,\mathbf{0}]^H$.

Note that \textbf{P5} is identical to \textbf{P2}, which is a constrained
regularized least square problem. Thus, \textbf{P5} has the same semi-closed-form
solution as provided in  Section \ref{Subsec:precoding} with complexity $\mathcal{O}(\min(N^c,K)^3)$
to compute the solution. At each time $t$, for each cell $c$,  based on the
inaccurate local CSI $\hat{\mathbf{H}}^c(t)$ and virtualization demand $\hat{\mathbf{D}}^c(t)$,
the InP obtains an optimal local precoding matrix $\hat{\mathbf{V}}^{c\star}(t)$
by solving $\textbf{P5}$, and then update the virtual queue $Z^c(t)$ according
to its queue dynamics in (\ref{eq:CVQ}). As such, the online per-slot coordinated
precoding problem \textbf{P4} leads to a fully-distributed implementation
at each cell, without any CSI exchange across cells. An outline of the proposed
multi-cell coordinated MIMO WNV algorithm is given in Algorithm~\ref{alg:2}.

\begin{algorithm}[t]
\begin{small}
\caption{Outline of Online Multi-Cell Coordinated  MIMO WNV Algorithm }\label{alg:2}
\begin{algorithmic}[1]
\STATE Set $U>0$ and $\mathbf{Z}(0)=\mathbf{0}$.
\STATE At each time $t$, obtain $\hat{\mathbf{H}}^c(t)$ and $Z^c(t)$ in each
cell $c$.
\STATE Solve  \textbf{P5} for $\hat{\mathbf{V}}^{c\star}(t),\forall{c}\in\mathcal{C}$
(see Section \ref{Subsec:precoding}).
\STATE Update  $Z^c(t+1) = \max\{Z^c(t)+ \Vert \hat{\mathbf{V}}^{c\star}(t)\Vert_F^2-\bar{P}^c,0\},\forall{c}\in\mathcal{C}$.
\end{algorithmic} 
\end{small}
\end{algorithm}

\subsection{Performance Bounds}

Similar to the single-cell case, for performance analysis, we assume that
the global channel gain $\Vert\mathbf{H}'(t)\Vert_F$ is bounded by a constant
$B>0$ for any $t$ as in (\ref{eq:asm1}). With given channel estimation quality
as in (\ref{eq:asm2}), we assume that the normalized CSI inaccuracy is bounded
by a constant $\delta\geq0$ at any $t$ as
\begin{align}
        \frac{\Vert\tilde{\mathbf{H}}_m^{cl}(t)\Vert_F}{\Vert\mathbf{H}_m^{cl}(t)\Vert_F}\leq\delta,\quad\forall{m}\in\mathcal{M},\quad\forall{c},l\in\mathcal{C},\quad\forall{t}\label{eq:Casm2}
\end{align}
where $\tilde{\mathbf{H}}_m^{cl}(t)\triangleq\mathbf{H}_m^{cl}(t) - \hat{\mathbf{H}}_m^{cl}(t)$
is the channel estimation error and $\hat{\mathbf{H}}_m^{cl}(t)$ is the estimated
channel state between SP $m$'s users in cell $c$ and the BS $l$.
It follows that the estimated channel gain $\Vert\hat{\mathbf{H}}'(t)\Vert_F$
is bounded by $B(1+\delta)$ as in (\ref{Hhat}), for any $t$.

Similar to  Lemma \ref{lmZ}, we show below that the virtual queue $Z^c(t)$
produced by Algorithm \ref{alg:2} is upper bounded at any $t$.
\begin{lem}\label{lmZC}
        By Algorithm \ref{alg:2}, $Z^c(t)$ satisfies
        \begin{align}
        Z^c(t)\leq U B^2(1+\delta)^2 \xi^c+P_{\text{max}}^c-\bar{P}^c,\quad\forall{c}\in\mathcal{C},\quad\forall{t}\label{CBDVQ}
        \end{align}
        where $\xi^c\triangleq\sqrt{\frac{N^c}{\bar{P}^c}\sum_{m\in\mathcal{M}}P_m^c},\forall{c}\in\mathcal{C}$.
\end{lem}
\textit{Proof:} See Appendix \ref{APP:lmZC}. \endIEEEproof

Let $\mathcal{M}_{\text{\begin{tiny}MRT\end{tiny}}}^c=\{1,\dots,M_{\text{\begin{tiny}MRT\end{tiny}}}^c\}$
be the set of SPs that adopt MRT precoding in cell $c$. Similar to (\ref{MRT})
and (\ref{ZF}), each  SP $m\in\mathcal{M}_{\text{\begin{tiny}MRT\end{tiny}}}^c$
and $m\in\mathcal{M}\backslash\mathcal{M}_{\text{\begin{tiny}MRT\end{tiny}}}^c$
uses the following MRT and ZF precoding, respectively:
\begin{align*}
        \hat{\mathbf{W}}_{m}^{\text{\begin{tiny}MRT\end{tiny}},cc}(t)&=\sqrt{P_m^c}\frac{\hat{\mathbf{H}}_{m}^{ccH}(t)}{\Vert\hat{\mathbf{H}}_{m}^{cc}(t)\Vert_F},\notag\\
        \quad\hat{\mathbf{W}}_{m}^{\text{\begin{tiny}ZF\end{tiny}},cc}(t)&=\sqrt{P_m^c}\frac{\hat{\mathbf{H}}_{m}^{ccH}(t)(\hat{\mathbf{H}}_{m}^{cc}(t)\hat{\mathbf{H}}_{m}^{ccH}(t))^{-1}}{\sqrt{\tr\{(\hat{\mathbf{H}}_{m}^{cc}(t)\hat{\mathbf{H}}_{m}^{ccH}(t))^{-1}\}}}
\end{align*}
where we assume $K_m^c\le{N}^c$. Similar to Lemma \ref{Dbd}, based on each
SP's precoding scheme, we show below that given the CSI inaccuracy $\delta$
in (\ref{eq:Casm2}), the deviation between the accurate and inaccurate virtualization
demands $\Vert\mathbf{D}'(t)-\hat{\mathbf{D}}'(t)\Vert_{F}$ is upper bounded
by $\mathcal{O}(\delta)$, for any time $t$.

Specifically, define $\hat{B}_m^{c,\text{\begin{tiny}min\end{tiny}}}\triangleq\min\{\Vert\hat{\mathbf{H}}_m^{cc}(t)\Vert_F:\forall{t}\}$.
Let $\hat{\omega}_m^{c,\text{\begin{tiny}min\end{tiny}}}$ and $\omega_m^{c,\text{\begin{tiny}min\end{tiny}}}$
be the minimum eigenvalues of $\hat{\mathbf{H}}_m^{cc}\hat{\mathbf{H}}_{m}^{ccH}$
and $\mathbf{H}_m^{cc}(t)\mathbf{H}_m^{ccH}(t)$, respectively, over all $t$'s.
They indicate the minimum channel gain of $\hat{\mathbf{H}}_m^{cc}(t)$, and
the minimum energy in the eigen-directions for both $\hat{\mathbf{H}}_m^{cc}(t)$
and $\mathbf{H}_m^{cc}(t)$. We have the following lemma.
\begin{lem}\label{CDbd}
At each time $t$, the following hold:
\begin{align}
        \Vert\mathbf{D}'(t)\Vert_F&\leq  \zeta' B, \notag\\
        \Vert\hat{\mathbf{D}}'(t)\Vert_F&\leq \zeta' B(1+\delta),\notag\\
        \Vert \mathbf{D}'(t)-\hat{\mathbf{D}}'(t) \Vert_F&\leq\eta'{B}\delta\label{CBDDDhat}
\end{align}
where $\eta'\triangleq\sqrt{\sum_{\displaystyle{c}\in\mathcal{C}}\left(\sum_{m\in\mathcal{M}_{\text{\begin{tiny}MRT\end{tiny}}}^c}\alpha_m^c+\sum_{m\in\mathcal{M}\backslash\mathcal{M}_{\text{\begin{tiny}MRT\end{tiny}}}^c}\beta_m^c\right)}$,
$\alpha_m^c\triangleq\left(1+\frac{(2+\delta)B}{\hat{B}_m^{c,\text{\begin{tiny}min\end{tiny}}}}\right)^{2}P_m^c$,
$\beta_m^c\triangleq\left(\frac{B^4(1+\delta)^2}{K_m^c\hat{\omega}_m^{c,\text{\begin{tiny}min\end{tiny}}}\omega_m^{c,\text{\begin{tiny}min\end{tiny}}}}\right)^2P_m^c$,
and $\zeta'\triangleq\sqrt{\sum_{c\in\mathcal{C}}\sum_{m\in\mathcal{M}}P_m^c}$.
\end{lem}
\textit{Proof:} See Appendix \ref{APP:CDbd}. \endIEEEproof

Define $\phi'(\mathbf{H}'(t),\mathbf{V}'(t),\mathbf{D}'(t))\!\triangleq\!{U}\Vert\mathbf{H}'(t)\mathbf{V}'(t)\!-\!\mathbf{D}'(t)\Vert_F^2\!+\!\sum_{c\in\mathcal{C}}Z^c(t)\Vert\mathbf{V}^c(t)\Vert_F^2$,
and note that $\phi'(\hat{\mathbf{H}}'(t),\hat{\mathbf{V}}'(t),\hat{\mathbf{D}}'(t))$
is the objective function in \textbf{P4}. Based on Lemma \ref{CDbd}, we show
in the following lemma that the performance gap between using the optimal
solution $\hat{\mathbf{V}}^{\prime\star}(t)$
to \textbf{P4} under the inaccurate channel state $\hat{\mathbf{H}}'(t)$
and using the optimal solution $\mathbf{V}^{\prime\text{\begin{tiny}opt\end{tiny}}}(t)$
to \textbf{P3} under the accurate channel state $\mathbf{H}'(t)$ is upper
bounded by $\mathcal{O}(\delta)$.

\begin{lem}\label{Clmrho}
At each time $t$, the following holds:     
\begin{align}
        \!\!\!\!\phi'(\mathbf{H}'(t),\!{\hbox{$\hat{\mathbf{V}}'$}}^{\star}(t),\!\mathbf{D}'(t))\!-\!\phi'(\mathbf{H}'(t),\!\mathbf{V}^{\prime\text{\begin{tiny}opt\end{tiny}}}(t),\!\mathbf{D}'(t))\!\leq\!{U}\varphi'\!\!\!\label{CP2BD}
\end{align}
where
\begin{align*}
        \varphi'\!\triangleq\!2\!\left[(2+\delta)({\gamma'}^2+\zeta'\eta')+2(\zeta'(1+\delta)+\eta')\gamma'\right]\!B^{2}\delta\!=\!O(\delta)
\end{align*}
with $\gamma'\triangleq\sqrt{\sum_{c\in\mathcal{C}}P_{\text{max}}^c}$.

\end{lem}
\textit{Proof:} The proof is similar to the proof of Lemma \ref{Dbd} by applying
(\ref{CBDDDhat}) in Lemma \ref{CDbd}, and hence is omitted.\endIEEEproof

Following Lemma \ref{Clmrho}, we have the following upper bound on the expected
DPP metric using the optimal precoding solution $\hat{\mathbf{V}}^{\prime\star}(t)$
to \textbf{P4}. 
\begin{lem} \label{lmdppC}
At each time $t$, we have
\begin{align}
        &\mathbb{E}\{\Delta(t) \} + U\mathbb{E}\{\Vert\mathbf{H}'(t)\hat{\mathbf{V}}^{\prime\star}(t)-\mathbf{D}'(t)\Vert_F^{2}\}\notag\\
        &\leq{U}\mathbb{E}\{\Vert\mathbf{H}'(t){\hbox{$\mathbf{V}'$}}^{\star}(t)-\mathbf{D}'(t)\Vert_F^2\}+U\varphi'+S'\label{Ceqdpp}
\end{align}
where $S'$ and $\varphi'$ are given in Lemma \ref{lmDPPUBC} and \ref{Clmrho},
respectively.
\end{lem}
\textit{Proof:} See Appendix \ref{APP:lmdppC}.\endIEEEproof

Finally, with Lemmas \ref{lmZC} and \ref{lmdppC}, we have the following performance
bounds for Algorithm \ref{alg:2} in the multi-cell scenario with imperfect
CSI over any given time horizon $T$.

\begin{thm}\label{lmCHBD}
Given any $\epsilon>0$, set $U=\frac{S'}{\epsilon}$ in Algorithm~\ref{alg:2}.
For any $T>0$, for $\hat{\mathbf{V}}^{\prime\star}(t)$ produced by Algorithm
\ref{alg:2} with $\hat{\mathbf{H}}'(t)$, the following hold regardless of
the distribution of $\mathbf{H}'(t)$:
\begin{align}
        &\!\!\!\!\frac{1}{T}\sum_{t=0}^{T-1} \mathbb{E}\left\{ \Vert\mathbf{H}'(t)\hat{\mathbf{V}}^{\prime\star}(t)-\mathbf{D}'(t)\Vert_F^2\right\}\!\leq\rho^{\prime\text{\begin{tiny}opt\end{tiny}}}+\varphi'+\epsilon,\label{Cthm1}\\
        &\!\!\!\!\frac{1}{T}\!\sum_{t=0}^{T-1} \Vert \hat{\mathbf{V}}^{c\star}(t)\Vert_F^2\leq\bar{P}^c+\frac{S'B^2(1\!+\!\delta)^2\xi^c\!+\!\epsilon(P_{\text{max}}^c\!-\!\bar{P}^c)}{\epsilon{T}}\!\!\label{Cthm2}
\end{align}
where $\rho^{\prime\text{\begin{tiny}opt\end{tiny}}}$ is the minimum objective
value of \textbf{P3} under $\mathbf{H}'(t)$, $\varphi'$ is defined below
(\ref{CP2BD}), and $\xi^c$ is defined below (\ref{CBDVQ}).    
\end{thm}
\textit{Proof:} See Appendix \ref{APP:lmCHBD}. \endIEEEproof

The upper bound in (\ref{Cthm1}) on the objective value of \textbf{P3} indicates
that, similar to Algorithm~\ref{alg:1} for the single-cell case, for any
given $T$, the performance of Algorithm \ref{alg:2} using $\hat{\mathbf{H}}'(t)$
for the multi-cell case can still be arbitrarily close to the optimum achieved
with true channel state $\mathbf{H}'(t)$ plus a gap of $O(\delta)$. Furthermore,
(\ref{Cthm2}) provides a bound on the per-cell time-averaged transmit power
for any given $T$. The bound indicates that for all $T\geq\frac{1}{\epsilon^2}$,
Algorithm \ref{alg:2} guarantees that the deviation from the long-term transmit
power limit $\bar{P}^c$ at each cell $c$ is within $O(\epsilon)$.

\section{Simulation Results}\label{Sec:Simulation Results}

In this section, we present our simulation studies under the typical urban
micro-cell LTE network settings. We study the values of the design parameters
in the proposed algorithm as well as the effect of various system parameters
on the performance.

\begin{figure}[!t]
\centering
\vspace{0mm} 
\subfloat[All SPs adopt MRT precoding.]
{\includegraphics[width=.9\linewidth,trim=00 00 00 00,clip]{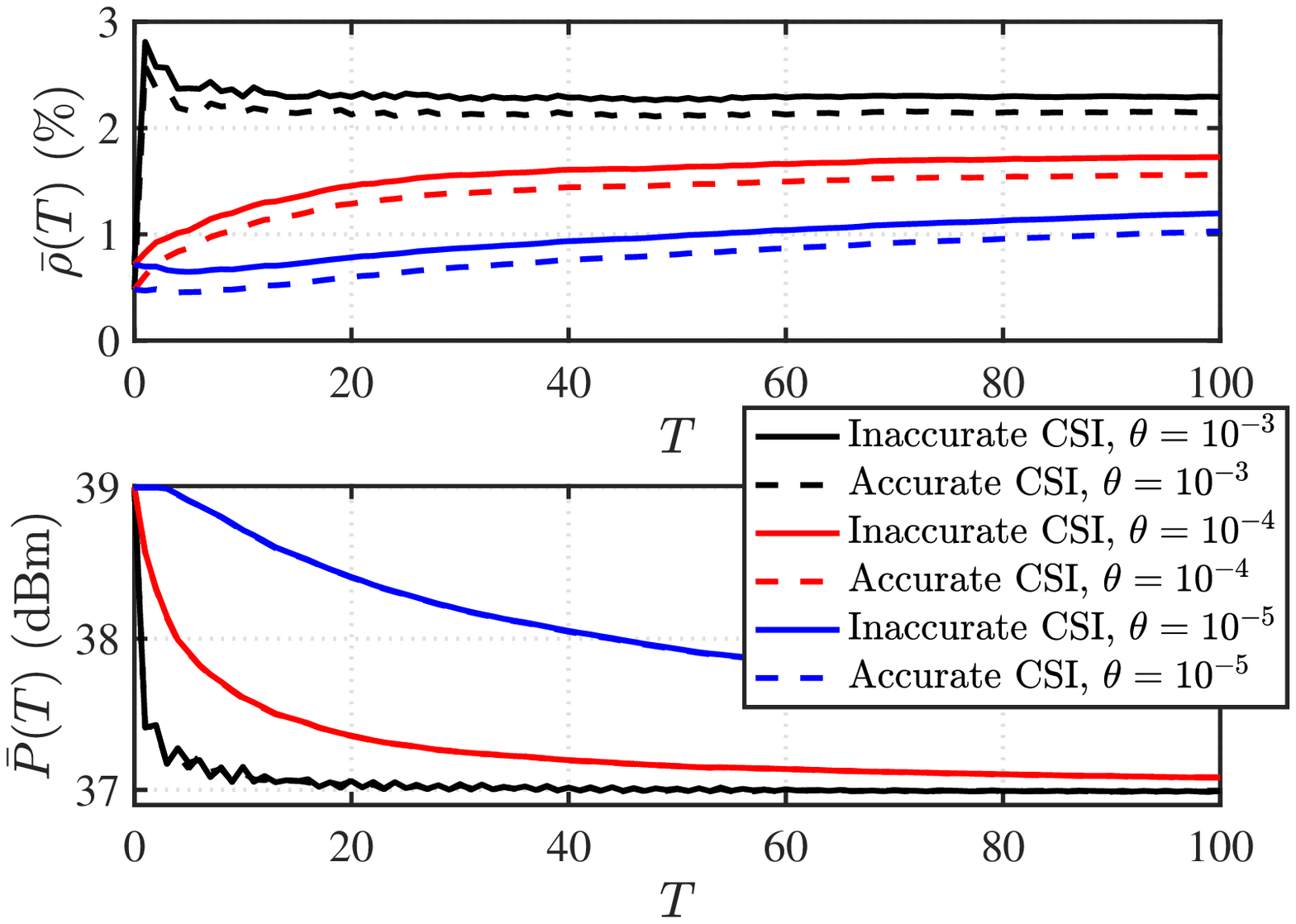}}
\vspace{0mm} 
\subfloat[All SPs adopt ZF precoding.]
{\includegraphics[width=.9\linewidth,trim=00 00 00 00,clip]{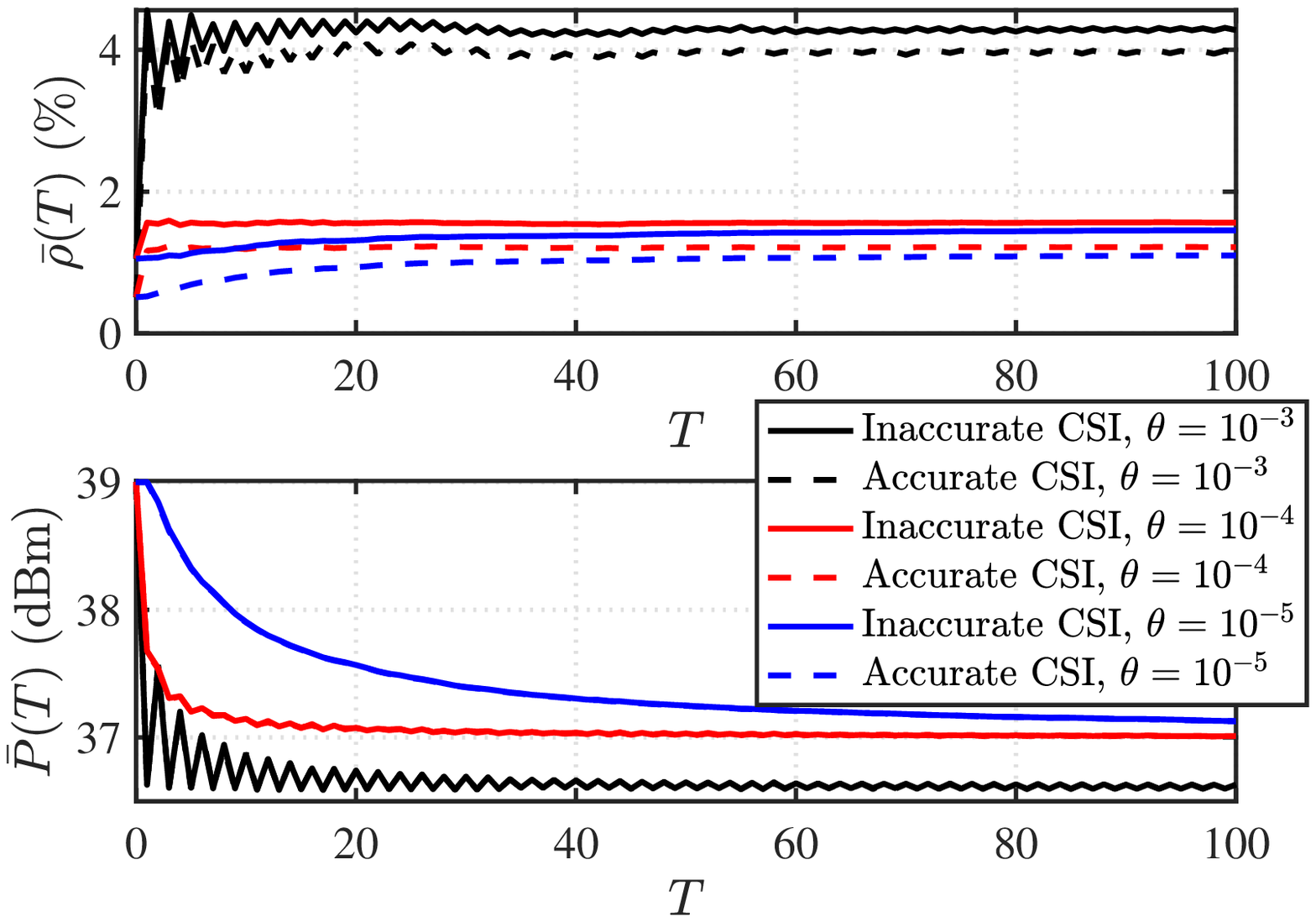}}
\caption{$\bar{\rho}(T)$ and $\bar{P}(T)$ vs. $T$ under different precoding
schemes adopted by SPs.}\label{fig:theta}
\vspace{0mm} 
\end{figure}

\begin{figure}[!t]
\centering
\vspace{0mm} 
\subfloat[All SPs adopt MRT precoding.]
{\includegraphics[width=.9\linewidth,trim=00 00 00 00,clip]{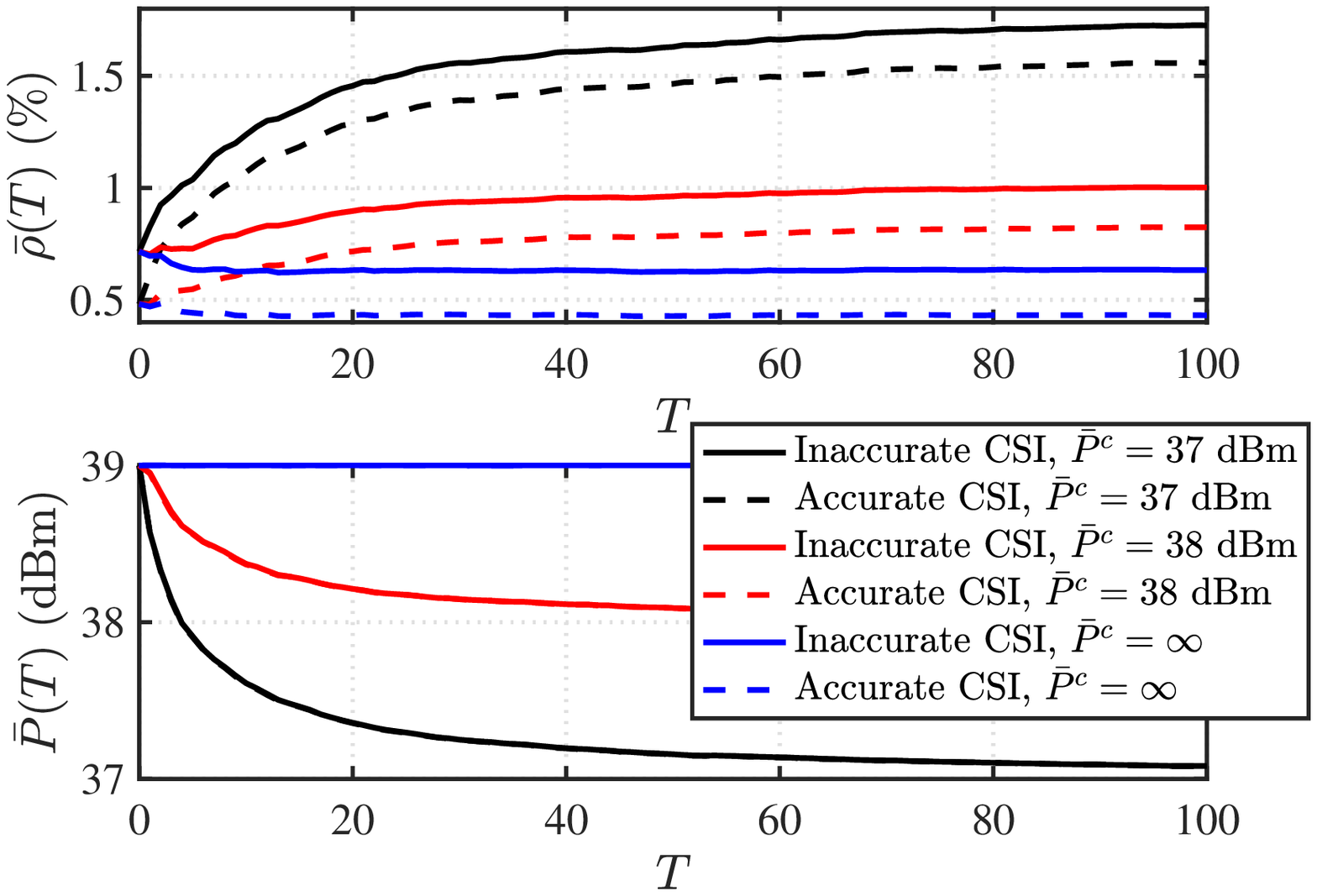}}
\vspace{0mm} 
\subfloat[All SPs adopt ZF precoding.]
{\includegraphics[width=.9\linewidth,trim=00 00 00 00,clip]{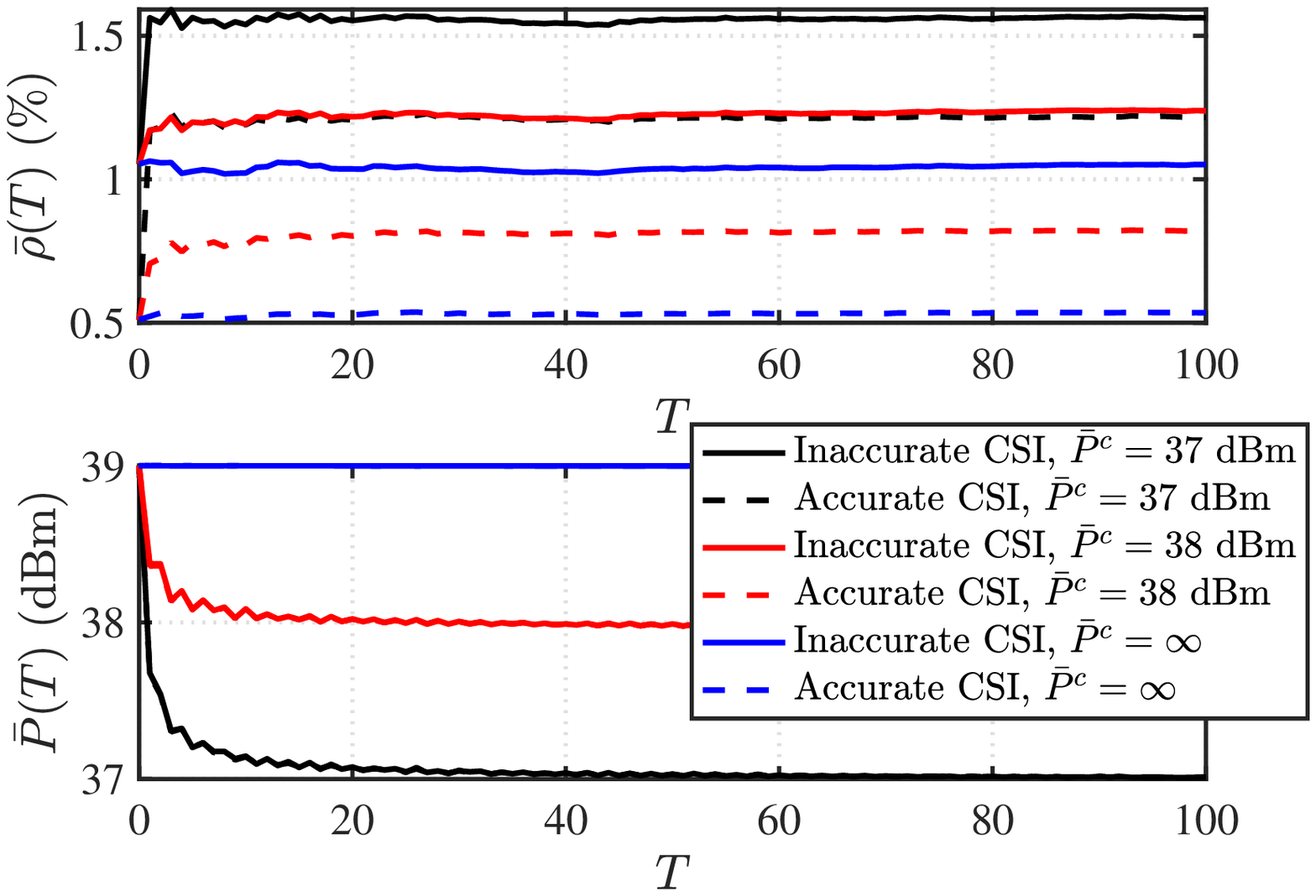}}
\caption{$\bar{\rho}(T)$ and $\bar{P}(T)$ vs. $T$ under different values
of $\bar{P}^c$.}\label{fig:barP}
\vspace{0mm} 
\end{figure}

\subsubsection{Simulation Setup}

We consider an InP that owns a virtualized network consisting of $C=7$ urban
hexagon micro cells, each with radius $R=500~m$. The InP-owned BS at the
center of cell $c\in\mathcal{C}$ is equipped with $N^c=32$ antennas. The
InP serves $M=4$ SPs. Each SP $m$ has $K_m^c=2$ subscribing users uniformly
distributed in  cell $c$. Following the typical LTE specifications~\cite{LTEP},
we focus on the channel over bandwidth $B_W=60$ kHz, which is the sum bandwidth
of $M$ subcarriers. We set the maximum transmit power limit over the channel
to be $P_{\text{max}}^c=39~\text{dBm},\forall{c}\in\mathcal{C}$. Unless it
is specified, we set the default average transmit power $\bar{P}^c=37$~dBm.
The receiver noise spectral density is $N_0=-174$ dBm/Hz, and the noise figure
is set to $N_F=10$~dB. At each time $t$, the channel from user $k$ of SP
$m$ in cell $c$ to BS~$l$ is modeled as $\mathbf{h}_{mk}^{cl}(t)=\sqrt{\beta_{mk}^{cl}}\mathbf{g}_{mk}^{cl}(t)$,
where $\mathbf{g}_{mk}^{cl}(t)\sim\mathcal{CN}(\mathbf{0},\mathbf{I})$, and
$\beta_{mk}^{cl}$ represents the large-scale variation. We model $\beta_{mk}^{cl}$
as \cite{LTEP} $\beta_{mk}^{cl}[\text{dB}]=-31.54-33\log_{10}(d_{mk}^{cl})+\psi_{mk}^{cl}$,
where $d_{mk}^{cl}$ is the distance from BS $l$ to user $k$ of SP $m$ in
cell $c$, and $\psi_{mk}^{cl}\sim\mathcal{CN}(0,\sigma_{\phi}^2)$ is the
shadowing with $\sigma_{\phi}=8$ dB. For a given channel $h_{mk}^{cln}$ from
antenna $n$ of BS $l$ to user $k$ of SP $m$ in cell~$c$, we denote $e_{\mathbf{H}}$
as the standard deviation of the normalized CSI error, \ie $\frac{\tilde{h}_{mk}^{cln}}{|h_{mk}^{cln}|}\sim\mathcal{CN}(0,e_\mathbf{H}^2)$.
Finally, we assume each channel is i.i.d. over time.

To study the performance of Algorithm \ref{alg:2}, we consider the following
two metrics: First, we define the $T$-slot normalized
time-averaged precoding deviation from virtualization demand as $\bar{\rho}(T)\triangleq\frac{1}{T}\sum_{t=0}^{T-1}\frac{\Vert\mathbf{H}'(t)\hat{\mathbf{V}}^{\prime\star}(t)-\mathbf{D}'(t)\Vert_F^2}{\Vert\mathbf{D}'(t)\Vert_F^2}$;
Second, we consider the $T$-slot time-averaged per-cell transmit power as
$\bar{P}(T)\triangleq\frac{1}{TC}\sum_{t=0}^{T-1}\Vert\hat{\mathbf{V}}^{\prime\star}(t)\Vert_F^2$.
We assume that the InP allocates the transmit power $P_m^c=\frac{P_{\text{max}}^c}{M}$
to each SP $m$, $\forall{c}\in\mathcal{C}$.

\subsubsection{Effect of Weight $U$}

Recall that weight $U$ is a design parameter in Algorithm~\ref{alg:2}. We
first study the effect of weight $U=\frac{S'}{\epsilon}$ on the performance
of the proposed algorithm by varying $\epsilon$. Note that $\bar{\rho}(T)$
is normalized to the network-wide demand $\Vert\mathbf{D}'(t)\Vert_F^{2}$.
We use the upper bound for $\Vert\mathbf{D}'(t)\Vert_F^2$ in Lemma~\ref{CDbd}
to set $\epsilon$ as $\epsilon=\theta{\zeta'}^2B^2$, where $\theta$ is
used as a controllable parameter. From~(\ref{eq:asm1}),
we set $B=1.645\sqrt{\sum_{c\in\mathcal{C}}N^c\sum_{k\in\mathcal{K}^c}\beta_k^c}$,
which ensures that $\mathbb{P}\{\Vert\mathbf{H}'(t)\Vert_F>B\}<4.9\times10^{-12}$
based on the Chernoff bound.

\begin{figure}[!t]
\centering
\vspace{0mm} 
\subfloat[All SPs adopt MRT precoding.]
{\includegraphics[width=.9\linewidth,trim=00 00 00 00,clip]{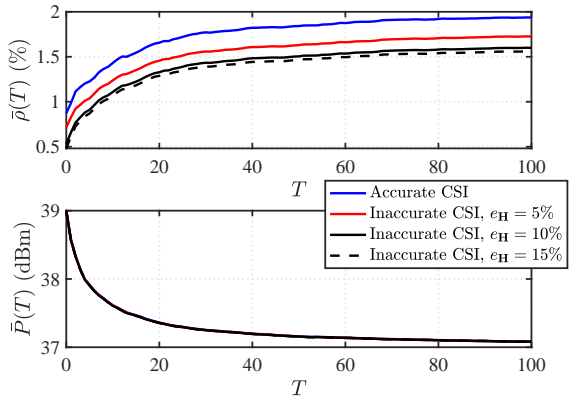}}
\vspace{0mm} 
\subfloat[All SPs adopt ZF precoding.]
{\includegraphics[width=.9\linewidth,trim=00 00 00 00,clip]{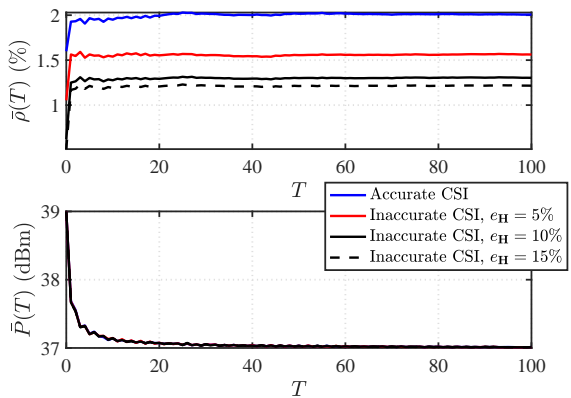}}
\caption{$\bar{\rho}(T)$ and $\bar{P}(T)$ vs. $T$ under different values
of $e_{\mathbf{H}}$.}\label{fig:pilot}
\vspace{0mm} 
\end{figure}

Figs.~2(a) and 2(b) show the time trajectory of $\bar{\rho}(T)$ and $\bar{P}(T)$
under different values of $\theta$, when all  SPs adopt MRT and ZF precoding,
respectively. The value of $\bar{\rho}(T)$ is shown in percentage. Both accurate
and inaccurate CSI are considered. We see that the $\bar{\rho}(T)$ under
inaccurate CSI closely follows that under perfect CSI. Since the power constraint
does not depend on the channels, as expected, $\bar{P}(T)$ under inaccurate
CSI is almost identical to that under perfect CSI. These performance in Fig.~\ref{fig:theta}
demonstrate that our proposed algorithm is robust to inaccurate CSI under
different precoding schemes adopted by SPs. Furthermore, we observe that
our proposed algorithm converges fast (within 100 time slots) for various
values of $\theta$. As $\theta$ decreases, $U$ becomes larger, which puts
more emphasis on the precoding deviation $\hat{\rho}'(t)$ than on the Lyapunov
drift $\Delta(t)$ in the DPP metric. As a result, it takes a longer time
for the virtual queue $\mathbf{Z}(t)$ to stabilize and for the performance
to reach the steady state. In addition, for a smaller value of $\theta$,
the steady state value of $\bar{\rho}(T)$ is smaller, and that of $\bar{P}(T)$
converges to $\bar{P}^c$. These behaviors are consistent with the bound analysis
in (\ref{Cthm1}) and (\ref{Cthm2}) in Theorem \ref{lmCHBD}. As we see, for
$\theta=10^{-4}$,  the average deviation $\bar{\rho}(T)$ is under $2\%$ for
both the MRT and ZF precoding cases. Based on this result, we set $\theta=10^{-4}$
as the default value for the rest of simulation.

\subsubsection{Effect of Long-Term Transmit Power Limit $\bar{P}^c$}

To study the effect of long-term average transmit power limit $\bar{P}^c$
in (\ref{CVM2}), we show in Fig.~\ref{fig:barP} the time trajectory of $\bar{\rho}(T)$
and $\bar{P}(T)$ under different values of $\bar{P}^c$, when all SPs adopt
either MRT or ZF precoding. When $\bar{P}^c=\infty$, the precoding design
in \textbf{P3} is only subject to the short-term transmit power constraint
(\ref{CVM3}). With inaccurate CSI, the steady-state value of $\bar{\rho}(T)$
is only around $0.7\%$ for MRT precoding  and $1\%$ for  ZF precoding. When
$\bar{P}^c$ decreases, although $\bar{\rho}(T)$ increases for both MRT and
ZF precoding schemes, their values remain small. For example, when $\bar{P}^c=37$
dBm, with inaccurate CSI, the steady-state value of $\bar{\rho}(T)$ is around
$2\%$ for both MRT and ZF precoding schemes. As we observe, there is a trade-off
between the steady-state value of $\bar{\rho}(T)$ and $\bar{P}^c$. The InP
can use this trade-off to balance the transmit power consumption and the
deviation of actual precoding from the virtualization demand.

\begin{figure}[!t]
\centering
\vspace{0mm} 
\subfloat[MRT precoding.]
{\includegraphics[width=.9\linewidth,trim=00 00 00 00,clip]{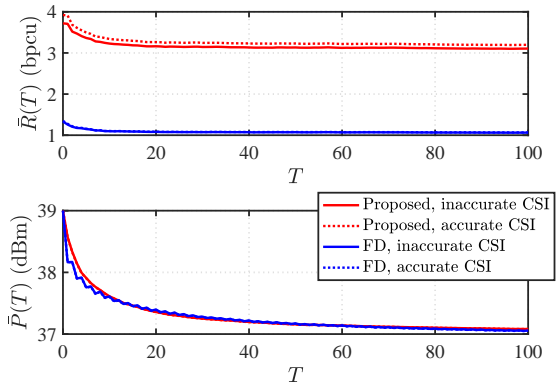}}
\vspace{0mm} 
\subfloat[ZF precoding.]
{\includegraphics[width=.9\linewidth,trim=00 00 00 00,clip]{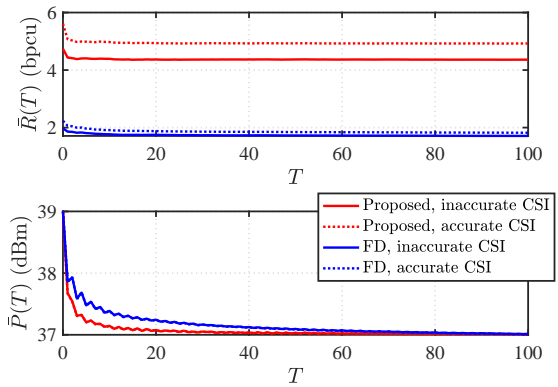}}
\caption{ Comparison of $\bar{R}(T)$ between the proposed approach
and FD approach.}\label{fig:NONV}
\vspace{0mm} 
\end{figure}

\subsubsection{Impact of Inaccurate CSI}

In Fig.~\ref{fig:pilot}, we study the impact of CSI inaccuracy on the performance
of the proposed algorithm by varying $e_\mathbf{H}$. As $e_{\mathbf{H}}$
increases from $5\%$ to $15\%$, the steady-state values of $\bar{\rho}(T)$
are still under $2\%$ for both MRT and ZF precoding schemes. Comparing the
two precoding schemes, we observe that $\bar{\rho}(T)$ is more sensitive
to $e_{\mathbf{H}}$ under ZF precoding than under MRT precoding. The reason
is that ZF precoding requires accurate CSI to null the inter-user interference,
and thus its performance is more sensitive to CSI accuracy \cite{MRTvsZF}.
In contrast, MRT precoding generates power gain at the general signal direction
and is less sensitive to the CSI accuracy. Finally, we observe that the steady-state
value of $\bar{P}(T)$ is similar for different values of $e_{\mathbf{H}}$,
showing that it is not sensitive to CSI accuracy.

\subsubsection{Benefit of Spatial Virtualization for Service Isolation}

Most existing works on MIMO WNV adopt the physical isolation approach to
separate the SPs  \cite{rp}-\nocite{EE17}\nocite{V5G}\nocite{CRAN}\nocite{WNVNOMA}\cite{Sgame}.
To the best of our knowledge, there is no existing online method for virtualization
in multi-cell MIMO systems.\footnote{For traditional
non-virtualized multi-cell systems, existing coordinated precoding schemes
focus on per-slot optimization
problems with per-slot maximum transmit power limit only. These per-slot
precoding solutions are
not comparable with the proposed online solution with the long-term transmit
power constraint.}  Therefore, for performance comparison, we implement a
physical isolation scheme for online multi-cell MIMO WNV. Specifically, we
consider a frequency division (FD) scheme that allocates equal bandwidth
$\frac{B_W}{M}$ to each SP $m$. We then use the proposed proposed online
coordinated precoding solution to serve each SP. For each SP, this can be
considered as a special case of Algorithm~\ref{alg:2} with a single SP, maximum
power limit $\frac{P_c^{\text{max}}}{M}$, and long-term power limit $\frac{\bar{P}_c}{M}$.

Fig. \ref{fig:NONV} shows the averaged user rate  $\bar{R}(T)\triangleq\frac{1}{TK}\!\sum_{t=0}^{T-1}\!\sum_{k\in\mathcal{K}}\log_2\left(\!1\!+\!\frac{|[\mathbf{H}'(t)\hat{\mathbf{V}}'(t)]_{k,k}|^2}{\sum_{j\neq{k}}\!|[\mathbf{H}'(t)\hat{\mathbf{V}}'(t)]_{k,j}|^2+\sigma_n^2}\!\right)\!$
achieved by the proposed  approach and the FD approach for both inaccurate
CSI and accurate CSI. Note that all rates are normalized by the total bandwidth
$B_W$. For both the MRT and ZF precoding cases, $\bar{R}(T)$ under both approaches
quickly converges to its steady state. The average rate achieved by the proposed
 spatial isolation approach is $2{\raise.17ex\hbox{$\scriptstyle\sim$}}3$
times higher that of the FD approach. This indicates substantial performance
advantage of the proposed spatial isolation approach over the physical isolation
approach for online virtualization in a multi-cell MIMO network.

\section{Conclusions}
\label{Sec:Conclusions}

In this paper, we have considered designing  online downlink MIMO WNV in
a multi-cell network with imperfect CSI, where the InP provides a precoding
solution based on the SPs' independent service demands. Assuming fading channels
and bounded CSI estimation error, we propose an online multi-cell coordinated
precoding algorithm aiming to minimize the long-term time-averaged precoding
deviation of the InP's actual precoding solution from the virtualization
demands  by the SPs. Our proposed algorithm only depends on the imperfect
CSI estimates currently available at the SPs and the InP, without the knowledge
of the channel distribution. Our online coordinated precoding solution is
fully distributed and in semi-closed form, which can be implemented at each
cell without any CSI exchange across cells. Our analysis on the performance
of the proposed online algorithm  takes into account the two-fold impact
of imperfect CSI on both the InP and the  SPs, and we observe an optimality
gap of $\mathcal{O}(\delta)$ over any time horizon due to CSI inaccuracy~$\delta$.
Simulation results demonstrate the effectiveness of the proposed algorithm
in both convergence rate and performance robustness to imperfect CSI, as
well as superior performance over the physical isolation approach.
\appendices

\section{Proof of Lemma \ref{lmDPPUBC}}\label{APP:lmDPPUBC}

Similar to the proof of (\ref{eq:lm1-1}) in Lemma \ref{lmDPPUB}. From the
virtual queue updating rule in (\ref{eq:CVQ}), the per-cell short-term power
constraints (\ref{CVM3}), and $L(t)=\frac{1}{2}\sum_{c\in\mathcal{C}}Z^{c2}(t)$,
we can show that
\begin{align}
       \Delta(t)&\le\frac{1}{2}\sum_{c\in\mathcal{C}}\max\{(P_{\text{max}}^c-\bar{P}^c)^2,\bar{P}^{c2}\}\notag\\
       &\quad+\sum_{c\in\mathcal{C}}Z^c(t)(\Vert\hat{\mathbf{V}}^c(t)\Vert_F^2-\bar{P}^c).\label{eq:lmDPPUP-1}
\end{align}
Taking the conditional expectation over the virtual queue $\mathbf{Z}(t)$
and adding $U\mathbb{E}\{\hat{\rho}'(t)|\mathbf{Z}(t)\}$ at both sides of
the above inequality, we complete the proof.\endIEEEproof

\section{Proof of Lemma \ref{lmZC}}\label{APP:lmZC}

Similar to the proof of (\ref{eq:lm2-2}) in Lemma \ref{lmZ}, we can show
that
\begin{align*}
        \Vert\hat{\mathbf{V}}^{c\star}\Vert_F\le\frac{U}{Z^c}B^2(1+\delta)^2\sqrt{N^c\sum_{m\in\mathcal{M}}P_m^c}.
\end{align*}
It follows that the sufficient condition for $Z^c(t)$ to ensure  $\Vert \hat{\mathbf{V}}^{c\star}(t)\Vert_F^2\leq\bar{P}^c$
for any time $t$ is $Z^c(t)\geq U B^2(1+\delta)^2\xi^c $. Further noting
that $Z^c(t+1)\leq{Z}^c(t)+P_{\text{max}}^c-\bar{P}^c$ from (\ref{eq:CVQ}),
we complete the proof.\endIEEEproof

\section{Proof of Lemma \ref{CDbd}}\label{APP:CDbd}

The proofs of the first two inequalities of (\ref{CBDDDhat}) follow from
$\Vert\hat{\mathbf{D}}'(t)\Vert_F^2=\sum_{c\in\mathcal{C}}\Vert\hat{\mathbf{D}}^c(t)\Vert_F^2$
and
\begin{align*}
        \Vert \hat{\mathbf{D}}^c(t)\Vert_F^2&\leq\sum_{m\in\mathcal{M}}\Vert\hat{\mathbf{H}}_m^{cc}(t)\Vert_F^2\Vert\hat{\mathbf{W}}_m^{cc}(t)\Vert_F^{2}\notag\\
        &\le{B}^2(1+\delta)^2\sum_{m\in\mathcal{M}}P_m^c.
\end{align*}
We now give the proof outline of the last inequality of (\ref{CBDDDhat}).
Similar to (\ref{DMRT})
and (\ref{DZF}) in the proof of Lemma \ref{Dbd}, we can show for any ${m}\in\tilde{\mathcal{M}}^c$
\begin{align*}
        \Vert\mathbf{H}_m^{cc}(t)\mathbf{W}_m^{cc\text{\begin{tiny}MRT\end{tiny}}}(t)-\hat{\mathbf{H}}_m^{cc}(t)\hat{\mathbf{W}}_{m}^{cc\text{\begin{tiny}MRT\end{tiny}}}(t)\Vert_F\le\sqrt{\alpha_m^c}B\delta
\end{align*}
and for any $m\in\mathcal{M}\backslash\breve{\mathcal{M}}^c$
\begin{align*}
        \Vert\mathbf{H}_m^{cc}(t)\mathbf{W}_m^{cc\text{\begin{tiny}ZF\end{tiny}}}(t)-\hat{\mathbf{H}}_m^{cc}(t)\hat{\mathbf{W}}_m^{cc\text{\begin{tiny}ZF\end{tiny}}}(t)\Vert_F\le\sqrt{\beta_m^c}B\delta.
\end{align*}
Applying the above two inequalities to
\begin{align*}
        &\Vert\mathbf{D}'(t)-\hat{\mathbf{D}}'(t)\Vert_F^2\notag\\
        &=\sum_{c\in\mathcal{C}}\sum_{m\in\mathcal{M}}\Vert\mathbf{H}_m^{cc}(t)\mathbf{W}_m^{cc}(t)-\hat{\mathbf{H}}_m^{cc}(t)\hat{\mathbf{W}}_m^{cc}(t)\Vert_F^2
\end{align*}
we complete the proof.\endIEEEproof

\section{Proof of Lemma \ref{lmdppC}}\label{APP:lmdppC}

Adding $U\Vert \mathbf{H}'(t)\hat{\mathbf{V}}^{\prime\star}(t)-\mathbf{D}'(t)\Vert_F^2$
to both sides of (\ref{eq:lmDPPUP-1}), taking expectations, noting (\ref{CP2BD})
in Lemma \ref{Clmrho}, and from the iterated law of expectation, we can show
that
\begin{align*}
        &\mathbb{E}\left\{\Delta(t) \right\} + U \mathbb{E}\{\Vert\mathbf{H}'(t)\hat{\mathbf{V}}^{\prime\star}(t)-\mathbf{D}'(t)\Vert_F^2\}\\
        &\le{U}\mathbb{E}\{\Vert\mathbf{H}'(t)\mathbf{V}^{\prime\text{\begin{tiny}opt\end{tiny}}}(t)-\mathbf{D}'(t)\Vert_F^2\}\notag\\
        &\quad+\mathbb{E}\left\{\mathbb{E}\left\{\sum_{c\in\mathcal{C}}Z^c(t)(\Vert\mathbf{V}^{c\text{\begin{tiny}opt\end{tiny}}}(t)\Vert_F^2\!-\!\bar{P}^c)|\mathbf{Z}(t)\right\}\right\}\!+\!U\varphi'\!+\!S'\!.
\end{align*}
We complete the proof by noting
\begin{align*}
        &\mathbb{E}\left\{\sum_{c\in\mathcal{C}}Z^c(t)(\Vert\mathbf{V}^{c\text{\begin{tiny}opt\end{tiny}}}(t)\Vert_F^2-\bar{P}^c)|\mathbf{Z}(t)\right\}\notag\\
        &=\sum_{c\in\mathcal{C}}Z^c(t)\mathbb{E}\{\Vert\mathbf{V}^{c\text{\begin{tiny}opt\end{tiny}}}(t)\Vert_F^2-\bar{P}^c\}\leq0
\end{align*}
which follows from $\mathbf{V}^{c\text{\begin{tiny}opt\end{tiny}}}(t)$ being
independent
of $\mathbf{Z}(t)$, $\mathbf{Z}(t)\succeq\mathbf{0}$, and $\mathbb{E}\{
\Vert\mathbf{V}^{c\text{\begin{tiny}opt\end{tiny}}}\Vert_F^2\}\leq\bar{P}^c$.
\endIEEEproof 

\section{Proof of Theorem \ref{lmCHBD}}\label{APP:lmCHBD}

Rearranging term of (\ref{Ceqdpp}), summing over time $t\in\{0,\dots,T-1\}$,
dividing by $UT$, and noting that $\sum_{t=0}^{T-1}\mathbb{E}\{\Delta(t)\}=\frac{1}{2}\mathbb{E}\{\Vert\mathbf{Z}(T)\Vert_2^2\}-\mathbb{E}\{\Vert\mathbf{Z}(0)\Vert_2^2\}$,
$\mathbf{Z}(0)=\mathbf{0}$,  $\mathbf{Z}(t)\succeq\mathbf{0},\forall{t}$,
and $U=\frac{S'}{\epsilon}$, we have (\ref{Cthm1}).

For any $c\in\mathcal{C}$, summing $\Vert\hat{\mathbf{V}}^{c\star}(t)\Vert_F^2\le\bar{P}^c+Z^c(t+1)-Z^c(t)$
over $t$ yields  $\frac{1}{T} \sum_{t=0}^{T-1} \Vert\hat{\mathbf{V}}^{c\star}(t)
\Vert_F^2\le\bar{P}^c+\frac{Z^c(T)}{T}$. Substituting the upper bound of
the virtual queue in (\ref{CBDVQ}) and $U=\frac{S'}{\epsilon}$ into the %
above
inequality, we have (\ref{Cthm2}).\endIEEEproof

\bibliographystyle{IEEEtran}
\bibliography{RefINFOCOM2020}

\end{document}